# Discovering interpretable low-dimensional dynamics using maximum entropy

Michael C. Chung[1,2], Tarran Mohan[1], Purushottam D. Dixit[2,3,4,*], and Juan Guan[1, †]

1College of Pharmacy, Division of Chemical Biology and Medicinal Chemistry,
University of Texas at Austin, Austin, TX 78712, USA
2Department of Biomedical Engineering, Yale University, New Haven, CT 06511, USA
3Systems Biology Institute, Yale University, West Haven, CT 06516 USA
4Program in Computational Biology and Biomedical Informatics, Yale University, New Haven, CT 06511, USA

*email: purushottam.dixit@yale.edu
†email: juanguan@utexas.edu

**Abstract**

Models (i.e., governing equations) are fundamental to science and engineering. Advances in data acquisition now make it possible to extract interpretable, low-dimensional descriptions from high-dimensional observations. However, existing approaches sacrifice either interpretability for reconstruction accuracy or infer symbolic dynamics without relating latent coordinates to physically meaningful observables. Here we present *Edwin* (maximum **e**ntropy-**d**riven compression **w**ith **i**nterpretable **n**onlinear model discovery), a unified framework that simultaneously performs dimensionality reduction using the dynamic maximum entropy (DME) principle and discovers sparse symbolic models governing latent dynamics, as well as the coupling between learned features and external metadata. We validate *Edwin* on diverse simulated systems, including stochastic diffusion, the Ornstein–Uhlenbeck process, self-assembling particles, spiking neural populations, and low-rank recurrent neural networks, as well as on a noisy experimental time series of aggregating RNA–liposome complexes. Across all systems, *Edwin* recovers low-dimensional symbolic models that are physically interpretable and generalize to unseen conditions. Together, these results establish *Edwin* as a powerful framework for inferring interpretable, low-dimensional dynamics directly from high-dimensional data.

## Introduction

Models lie at the heart of scientific understanding, providing interpretable and accurate descriptions of complex phenomena with predictive power. Virtually every scientific discipline, from single-cell biology [1, 2] and neuroscience [3, 4] to materials engineering [5], is now awash in high-dimensional, time-resolved data. Although these datasets are rich, their volume and complexity can obscure the dynamical mechanisms that drive system-level behavior. Notably, recent evidence suggests that many such complex systems exhibit intrinsic low dimensionality [6]; that is, high-dimensional observations can often be explained by dynamics evolving on a much lower-dimensional manifold. This observation presents an opportunity to develop algorithms that extract interpretable low-dimensional latent models directly from high-dimensional data.

Ideally, an interpretable low-dimensional dynamical model should satisfy two complementary criteria. First, the latent variables that compress high-dimensional observations should correspond to meaningful system attributes: for example, quantities reflected in external metadata or experimentally measurable features. Second, the temporal evolution of these variables should be governed by explicit symbolic equations that reveal the underlying mechanisms. Together, these properties enable both semantic interpretability of the state space and mechanistic interpretability of the dynamics.

However, existing tools typically satisfy only one of these criteria. Dimensionality-reduction methods, such as principal component analysis or deep autoencoders, prioritize reconstruction accuracy and often produce latent variables that are difficult to relate to physical quantities or external metadata. Conversely, model-discovery techniques, including symbolic [7, 8, 9, 10, 11, 12] and sparse regression methods [13, 14, 15, 16], can recover governing equations from data but typically assume that the relevant state variables are known and do not intrinsically perform dimensionality reduction. Hybrid approaches that rely on neural networks to combine representation learning with equation discovery attempt to address both tasks simultaneously, but are typically highly parameterized, rely on nonconvex optimization, impose strong prior assumptions on the latent space [17, 18, 19], or produce latent coordinates whose physical meaning is difficult to establish [20, 21, 22]. Moreover, many real-world signals (such as concentration fields, fluorescence intensities, or probability distributions) are intrinsically nonnegative and structured, rendering standard techniques based on signed latent variables (e.g., PCA) poorly suited to these domains. Addressing these challenges requires a method that simultaneously identifies a meaningful low-dimensional latent representation and discovers concise, interpretable equations governing its evolution.

In this work, to address these challenges, we introduce *Edwin* (maximum **e**ntropy-**d**riven compression **w**ith **i**nterpretable **n**onlinear model discovery, in honor of E. T. Jaynes) (Fig. 1). *Edwin* compresses high-dimensional, nonnegative observations into two low-dimensional latent representations: time-dependent variables that capture system dynamics and feature-dependent variables that encode spatial or structural variation. Simultaneously, *Edwin* discovers sparse symbolic models governing both the evolution of the time-dependent latents and the coupling between feature-dependent latents and external metadata; for example, in a spatiotemporal system, this metadata may correspond to spatial coordinates. To achieve this, *Edwin* combines the dynamic maximum entropy (DME) principle [23, 24] for flexible, constraint-based compression with a library-based sparse regression approach for automated equation discovery. Notably, model complexity and dimensionality are inferred directly from the data.

We validate *Edwin* on a diverse set of simulated and experimental systems spanning stochastic, nonlinear, and biological dynamics. Simulated examples include stochastic diffusion, the Ornstein–Uhlenbeck process, self-assembling particle systems, and neural population spiking. We further consider two challenging settings: low-rank recurrent neural networks with known latent structure, where recovery of the correct dynamical subspace is verified using canonical correlation analysis, and a noisy experimental time series of aggregating RNA–liposome complexes obtained from fluorescence microscopy, where no ground-truth model is available. Across all cases, *Edwin* reconstructs effective low-dimensional dynamics from high-dimensional observations, identifies sparse symbolic equations directly from noisy data, and predicts system behavior under previously unseen conditions. By exploiting structure inherent in nonnegative, time-resolved signals, the framework provides a practical route to interpretable data-driven modeling.

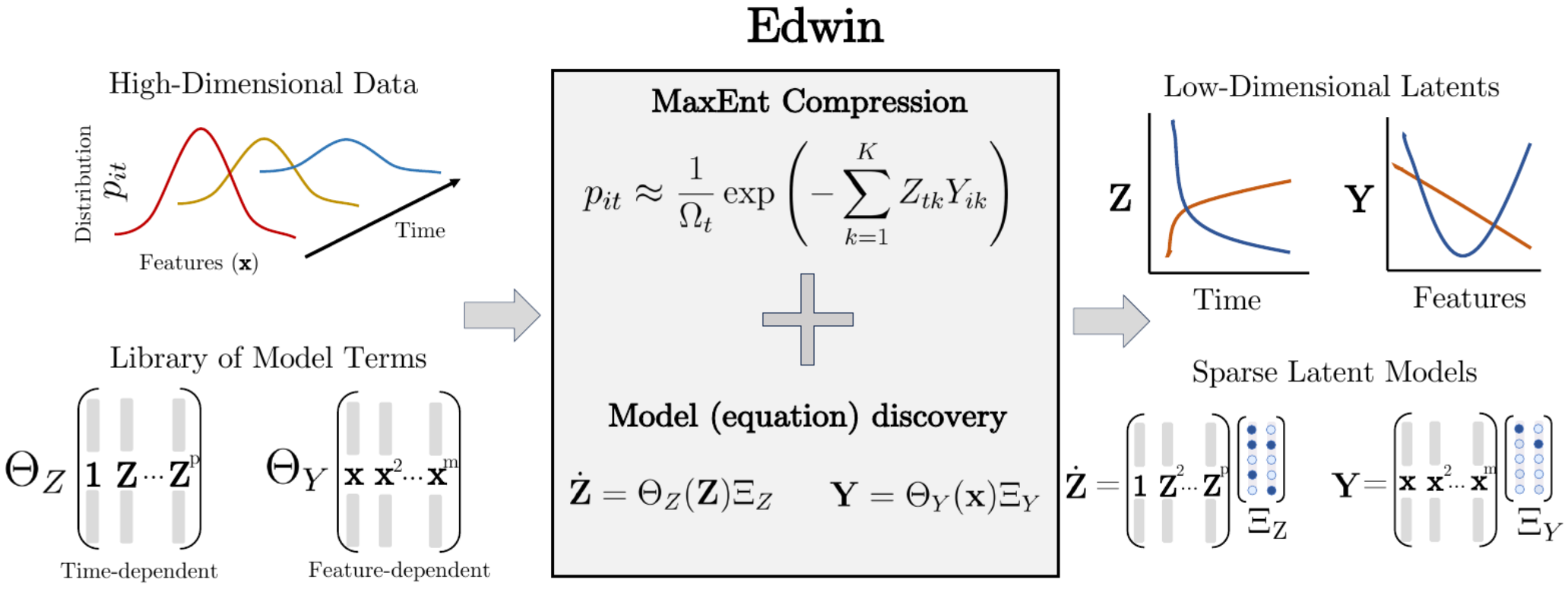

**Figure 1: Algorithmic overview of *Edwin*. *Edwin* takes as input nonnegative, normalized high-dimensional data, $p_{it}$, and libraries of model terms, $\Theta_Z$ and $\Theta_Y$. Here x denotes external feature metadata associated with each index $i$ (e.g., spatial coordinates for a spatiotemporal system, or cluster size for a self-assembly system). Using Dynamic Maximum Entropy (DME) for compression and sparse regression for model discovery, *Edwin* returns low-dimensional latents, Z, Y, and sparse models, $\Xi_Z$ and $\Xi_Y$, for the dynamics and features, respectively.**

## Background

In this section, we briefly describe the mathematical ingredients needed to formulate *Edwin*.

### *Dynamic Maximum Entropy*

Entropy maximization originates in the statistical mechanics developed by Gibbs, Maxwell, and Boltzmann [25, 26, 27], where it provides a principled method for inferring probabilities of microscopic configurations from incomplete macroscopic information. In this framework, microstates (such as the phase-space coordinates of particles) are assigned probabilities by selecting the distribution consistent with known ensemble averages (e.g., energy) that otherwise introduces no unwarranted assumptions.

A key conceptual advance by E. T. Jaynes was to recognize entropy maximization as a general principle for inference applicable across scientific domains [25, 26, 27]. In this formulation, one maximizes the entropy

$$S[p] = -\sum_i p_i \log p_i \tag{1}$$

subject to constraints expressed as expectation values $\langle Y_k \rangle = \sum_i p_i Y_{ik}$ for $k \in [1, K]$. The resulting distribution takes the Gibbs–Boltzmann form

$$p_i = \frac{1}{\Omega} \exp(-\sum_k Z_k Y_{ik}), \tag{2}$$

where $\Omega = \sum_i \exp(-\sum_{k=1}^{K} Z_k Y_{ik})$ is the partition function.

Viewing entropy maximization as an inference principle greatly expands its applicability beyond thermodynamics. By selecting the least biased distribution consistent with known constraints, it enables principled modeling from incomplete data and has been applied across diverse fields, including data science [28, 29], biology [30], epidemiology [31], and economics [32]. Notably, the Gibbs–Boltzmann distribution can also be interpreted as a form of dimensionality reduction [28], in which an enormous number of microscopic degrees of freedom (on the order of $10^{23}$, for physical systems) are effectively summarized by a small set of macroscopic variables. The maximum entropy principle thus provides a principled, theoretically grounded basis for inference, making maximally unbiased predictions consistent with known constraints [25]. In the present work, we exploit this foundation to simultaneously perform dimensionality reduction and model discovery directly from high-dimensional data.

Because our focus is on dynamical systems, Eq. (2) must be extended to time-dependent processes. One such extension is the dynamic maximum entropy (DME) approximation,

$$q_{it} = \frac{1}{\Omega_t}\exp(-\sum_{k=1}^{K} Z_{tk}\, Y_{ik}), \tag{3}$$

Here, in addition to the feature index $i \in [1, N]$, a time index $t \in [1, T]$ is introduced, and the time-dependent partition function $\Omega_t$ ensures normalization at each time point. In this work, this functional form is used to compress a time-resolved data matrix $p_{it}$ into a low-dimensional representation.

The DME approximation is well motivated both theoretically and empirically. For processes governed by the Fokker–Planck equation, DME satisfies the same H-theorem as the true dynamics, ensuring consistency with fundamental principles of nonequilibrium statistical mechanics [23]. It has also been applied successfully across a range of systems, including population dynamics [24, 33], flocking phenomena [34, 35], neural activity [36, 37], microbiome dynamics [38], and cosmic ray transport [39]. Crucially, however, these studies assume that either the underlying dynamical model or the structure of the latent variables is known *a priori*. In contrast, the present work infers both directly from data.

DME is closely related to other information-theoretic principles for inferring dynamics. The Principle of Maximum Caliber (MaxCal) [27, 26] extends entropy maximization to ensembles of trajectories, selecting the least biased distribution over paths consistent with dynamical constraints. From this perspective, DME can be viewed as a reduced description that focuses on time-dependent state distributions rather than full trajectory ensembles; Appendix 10 shows that DME arises as a special case of MaxCal under appropriate assumptions.

Another related framework is the Information Bottleneck Principle (IBP) [40, 41], which seeks compressed representations that preserve information relevant for prediction. In time-dependent settings, IBP favors representations that retain features informative about future states [42, 43]. For Markov processes governed by Fokker–Planck dynamics, these predictive features correspond to leading eigenfunctions of the transfer operator, that is, the slow modes governing long-timescale behavior [43], and the optimal encoding distribution takes an exponential-family form. DME produces structurally similar representations, but the underlying objectives differ: IBP explicitly optimizes predictive information, whereas DME enforces consistency with observed moment constraints at each time point. The two approaches coincide only in regimes where the imposed constraints align with the predictive slow modes.

### *A sparse regression algorithm for model discovery*

Library-based sparse regression has emerged as a powerful approach for constructing interpretable governing equations directly from data [13, 15]. The central idea is to discover parsimonious dynamical models by selecting a small number of active terms from a predefined library of candidate functions. Below, we briefly describe the formulation used in this work.

Let $\mathbf{X}, \mathbf{Y} \in \mathbb{R}^{n\times m}$ denote input–output matrix pairs, where $n$ is the number of data samples and $m$ is the number of state variables. Our goal is to identify an interpretable functional relationship between them. To this end, we construct a library matrix $\Theta(\mathbf{X}) \in \mathbb{R}^{n\times l}$ whose columns consist of user-specified nonlinear functions of $\mathbf{X}$; here $l$ is the number of candidate functions. The task is then to determine a coefficient matrix $\Xi \in \mathbb{R}^{l\times m}$ such that

$$\mathbf{Y} = \Theta(\mathbf{X})\Xi + \eta, \tag{4}$$

where $\eta \in \mathbb{R}^{n\times m}$ represents noise, assumed to be Gaussian with zero mean and unknown variance. Each column of $\mathbf{Y}$ is thus expressed as a linear combination of candidate nonlinear functions, with coefficients given by the corresponding column of $\Xi$. This formulation corresponds to generalized nonlinear regression [44] and provides interpretability through an explicit symbolic relationship between inputs and outputs.

When the objective is to infer differential equations governing dynamics, the output matrix corresponds to time derivatives of the state, $\mathbf{Y} = \dot{\mathbf{X}}$. Estimating these derivatives numerically from noisy data is often challenging, as differentiation amplifies measurement noise. In this work, we circumvent this difficulty using a Galerkin projection approach, which employs a weak formulation of Eq. [y&x] together with integration by parts to avoid explicit numerical differentiation [15]. Details are provided in Appendix 8.4.

To determine the coefficient matrix $\Xi$, we solve the regularized optimization problem

$$\| \mathbf{Y} - \Theta(\mathbf{X})\Xi \|_2^2 + \nu^2 R(\Xi), \tag{5}$$

where $R(\Xi)$ promotes sparsity and $\nu$ controls the strength of regularization. In this work we use $R(\Xi) = \| \Xi \|_0$, which counts the number of nonzero entries in $\Xi$. Larger values of $\nu$ encourage sparser models by penalizing small coefficients more strongly; in practice, $\nu$ acts as a pruning threshold that removes terms whose magnitude falls below a specified level, making careful selection of this parameter essential.

We determine $\Xi$ automatically without requiring the user to specify $\nu$, as described in Appendix 8.1. Briefly, an initial estimate of $\Xi$ is obtained, after which a sequence of threshold values is applied to prune small or spurious coefficients. Each threshold yields a candidate sparse model characterized by a corresponding model complexity $k$.

### *A novel information-theoretic criterion for model selection*

We briefly describe a recently developed information-theoretic criterion for model selection in sparse regression: the sample-length-scaling logarithmic information criterion (SLIC). In this framework, the complexity-penalized likelihood takes the form

$$\text{SLIC} = -2\log L(\hat{\Xi}) + n\log(k), \tag{6}$$

where $L(\hat{\Xi})$ is the maximum likelihood of the data under the sparsified model, $n$ is the number of samples, and $k$ is the model complexity (e.g., the number of active terms). Assuming independent Gaussian noise, $-2\log L(\hat{\Xi}) = n\log(\hat{\epsilon})$, where $\hat{\epsilon}$ denotes the mean-squared error [45], yielding SLIC $= n\log(\hat{\epsilon}k)$. For comparison, the Akaike and Bayesian information criteria (AIC and BIC) take the forms $n\log(\hat{\epsilon}) + 2k$ and $n\log(\hat{\epsilon}) + (\log n)k$, respectively [45].

Candidate models are ranked according to their SLIC scores, and the model with the lowest score is selected. This model is then used to generate a refined set of threshold values for sparsification, and the procedure is iterated to convergence. A toy example illustrating this pruning and selection process is provided in Appendix 8.2.

As discussed in Appendix 8.3, standard criteria such as AIC and BIC have a known deficiency in large-sample regimes. As $n \to \infty$, the goodness-of-fit term $n\log(\hat{\epsilon})$ dominates their sparsity penalties ($2k$ for AIC and $k\log n$ for BIC), which grow sublinearly with $n$. Consequently, these criteria tend to favor increasingly complex models as more data become available. SLIC avoids this behavior by employing a complexity penalty that scales as $n\log(k)$, maintaining a balance between fit quality and parsimony even for large datasets.

To evaluate the effectiveness of SLIC for model discovery, we simulated several nonlinear dynamical systems (Appendix 13) across a range of noise levels and trajectory lengths. As shown in Fig. 2, SLIC consistently recovers the correct governing equations under these diverse conditions. Having established its accuracy for model selection, we next integrate SLIC with DME compression to construct our joint framework.

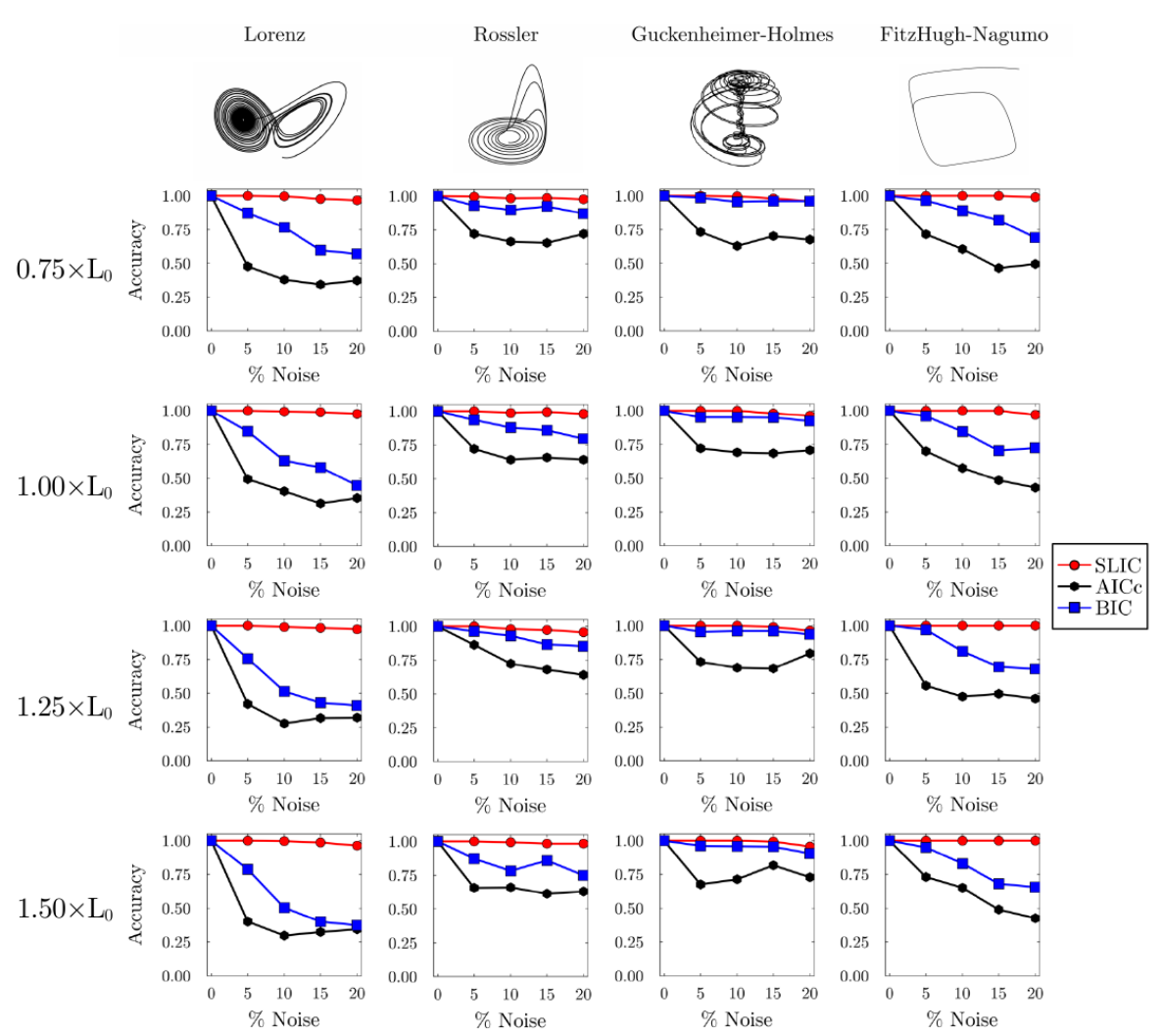

**Figure 2: Performance of the SLIC-Sparse Regression algorithm (Algorithm 1 in Appendix 8.1) in correctly recovering the governing equations of several nonlinear dynamical systems from data. The y-axis of each subplot is the accuracy (in the sense of a classification task) of the recovered model; 0 means that none of the correct terms are identified (no true positives or negatives) and 1 means the model is perfectly identified (no false positives or negatives). Going down the rows denotes a change in the trajectory length of the data, with $L_0$ denoting the default trajectory length, detailed in Appendix 13.**

## EDWIN: **e**ntropy-**d**riven compression **w**ith **i**nterpretable **n**onlinear model discovery

In this section, we integrate the ingredients developed above to formulate *Edwin*, a unified framework for maximum entropy–based discovery of low-dimensional dynamics.

Let $\mathbf{P} \in \mathbb{R}^{N\times T}$ be a positively valued data matrix with entries $p_{it}$, where $T$ denotes the number of time observations and $N$ the dimension of the feature space. Our objective is twofold. First, we compress $\mathbf{P}$ using the DME form in Eq. (3), obtaining latent factors $\mathbf{Z} \in \mathbb{R}^{T\times K}$ and $\mathbf{Y} \in \mathbb{R}^{N\times K}$ with elements $Z_{tk}$ and $Y_{ik}$, respectively, where $K$ is the latent dimension. Second, we infer models for both $\mathbf{Z}$ and $\mathbf{Y}$. Specifically, given libraries of candidate functions $\Theta_Z$ and $\Theta_Y$, we seek sparse coefficient matrices $\Xi_Z$ and $\Xi_Y$ satisfying

$$\dot{\mathbf{Z}} = \Theta_Z(\mathbf{Z})\Xi_Z \tag{7}$$

$$\mathbf{Y} = \Theta_Y(\mathbf{x})\Xi_Y \tag{8}$$

To achieve these objectives, we construct and minimize the following loss function, $\mathcal{L} = \mathcal{L}(\mathbf{Z}, \mathbf{Y}, \Xi_Z, \Xi_Y)$:

$$\mathcal{L} = \mathcal{L}_{KLD}(\mathbf{Z}, \mathbf{Y}) + \lambda_{\dot{Z}}\mathcal{L}_{\dot{Z}}(\mathbf{Z}, \Xi_Z) + \lambda_Y\mathcal{L}_Y(\mathbf{Y}, \Xi_Y) \tag{9}$$

The first term on the right hand side (RHS), $\mathcal{L}_{KLD}$, is the Kullback-Leibler Divergence between our input data and our compression scheme:

$$\begin{aligned}\mathcal{L}_{KLD}(\mathbf{Z},\mathbf{Y}) &= \sum_{i,t} p_{it} \log\left(\frac{p_{it}}{q_{it}}\right) \\ &= \sum_{i,t,k} p_{it}\, Z_{tk} Y_{ik} + \sum_t \log \Omega_t\end{aligned} \tag{10,11}$$

In Eq. (11), we have used Eq. (3) and dropped the additive constant $\sum_{i,t} p_{it} \log p_{it}$. Minimizing $\mathcal{L}_{KLD}$ enforces accurate reconstruction of $\mathbf{P}$. The second and third terms on the RHS of Eq. (9) enforce Eqs. (7) and (8), respectively; $\lambda_{\dot{Z}}$ and $\lambda_Y$ are hyperparameters that vary the strength of these terms. These loss terms are:

$$\mathcal{L}_{\dot{Z}}(\mathbf{Z}, \Xi_Z) = \frac{1}{2}||\dot{\mathbf{Z}} - \Theta_Z(\mathbf{Z})\Xi_Z||_2^2 + \nu_Z^2||\Xi_Z||_0 \tag{12}$$

and

$$\mathcal{L}_{Y}(\mathbf{Y}, \Xi_Y) = \frac{1}{2}||\mathbf{Y} - \Theta_Y(\mathbf{x})\Xi_Y||_2^2 + \nu_Y^2||\Xi_Y||_0 \tag{13}$$

where $\mathbf{x}$ denotes external metadata. As noted previously, this metadata may correspond, for example, to spatial coordinates in a spatiotemporal system or to cluster size in the self-assembly system described in Sec. 4.7.

We minimize Eq. (9) using a joint gradient-descent–regression scheme. At each iteration, $\mathbf{Z}$ and $\mathbf{Y}$ are updated by gradient descent, while the nonzero entries of $\Xi_Z$ and $\Xi_Y$ are updated by regression, which exactly minimizes the quadratic terms in Eqs. (12) and (13). Every $M$ iterations, we sparsify $\Xi_Z$ and $\Xi_Y$ using the SLIC-based procedure described in Appendix 8.1. The regularization hyperparameters $\lambda_{\dot{Z}}$ and $\lambda_Y$ are selected via grid search; details are given in Appendix 9.2.

Several aspects of this optimization problem are worth emphasizing. First, the problem of minimizing $\mathcal{L}_{KLD}$ is convex under an alternating scheme that fixes either $\mathbf{Z}$ or $\mathbf{Y}$ and solves for the other [28]. In addition, strong convergence results are available for the sparse regression subproblem [46]. Although we do not adopt an alternating optimization strategy here, we show in Sec. 4.3 that multiple random initializations converge to the same latent structures and model forms, indicating robustness of the solution. This contrasts with conventional deep-learning approaches, whose objectives are fully nonconvex in all parameters and often yield non-identifiable latent representations [47].

Second, symbolic regression methods, including evolutionary approaches such as AI-Feynman [12] and PySR [8] and probabilistic search-based approaches such as the Bayesian Machine Scientist [48], all perform search over large, open symbolic expression spaces and can in principle represent highly general functional forms. SLIC-based sparse regression differs fundamentally from this family of methods: rather than searching over expression space, it performs information-theoretic model selection within a fixed, differentiable candidate library. This has three practical consequences. First, it avoids the combinatorial complexity of expression-space search, enabling substantially lower computational cost when the governing dynamics lie within the chosen library; empirically, SLIC is approximately three to four orders of magnitude faster than PySR across all systems tested (Fig. S2). Second, because SLIC operates within a differentiable framework, it integrates naturally into the joint DME optimization, whereas search-based symbolic regression produces discrete expressions that do not permit gradient-based joint optimization. Third, while BMS approximates Bayesian model evidence through MCMC sampling over an open hypothesis space, SLIC achieves the same conceptual goal, namely penalizing model complexity while rewarding goodness of fit, via a closed-form information criterion, and does so on directly observed variables. Our framework extends this by performing model selection in a latent coordinate system that is itself jointly learned from high-dimensional data, a problem that search-based symbolic regression is not designed to address.

Finally, neural-network-based methods that extract symbolic models from latent dynamics [22, 20] recover governing equations but cannot assign physical meaning to the latent coordinates themselves: the variables $\mathbf{Z}$ remain physically uninterpreted without additional post-hoc interrogation of the network, limiting one's ability to connect the learned dynamics to observable quantities of interest. Our approach resolves this by jointly learning, alongside the latent dynamics, a feature-dependent model $\mathbf{Y} = \Theta_Y(\mathbf{x})\Xi_Y$ that explicitly couples each latent coordinate to external metadata $\mathbf{x}$. Consider a Gaussian distribution evolving under purely diffusive dynamics in $(1+1)$ dimensions, $p(x,t) \propto \exp(-x^2/t)$. Existing methods may recover latent dynamics of the form $\dot{Z} \propto -Z^2$, but identifying $Z$ as the inverse variance requires prior knowledge of the system and further interrogation of the network. In contrast, *Edwin* simultaneously learns $\dot{Z} \propto -Z^2$ and the feature-dependent model $Y(x) \propto x^2$, yielding the approximation $p(x,t) \propto \exp(-Z(t)Y(x))$. The coupling of these two models immediately and unambiguously identifies $Z(t)$ as the inverse variance of the distribution, without any additional analysis.

## Results

We evaluate *Edwin* across several representative systems and compare its performance with related approaches. For each system, we briefly describe the data, assess reconstruction accuracy, examine the inferred latent models, and demonstrate predictive performance on unseen conditions.

### *Brief description of the studied systems*

We consider four diverse systems (Fig. 3) to evaluate the performance of *Edwin*: an ensemble of two-dimensional Brownian particles (2D Brownian), the stochastic Ornstein–Uhlenbeck (OU) process, a system of self-assembling particles, and a population of spiking neurons. Brief descriptions are provided below; additional details appear in Appendix 7.

The first two systems (2D Brownian and OU; Fig. 3A,B) are canonical stochastic processes extensively studied in stochastic thermodynamics [49]. For each system, we generate ensembles of stochastic trajectories and estimate the corresponding time-dependent probability densities using kernel methods [50]; these noisy density estimates serve as input to *Edwin*. The features (denoted $\mathbf{x}$) correspond to spatial grid points: two-dimensional coordinates for the Brownian system and one-dimensional positions for the OU process.

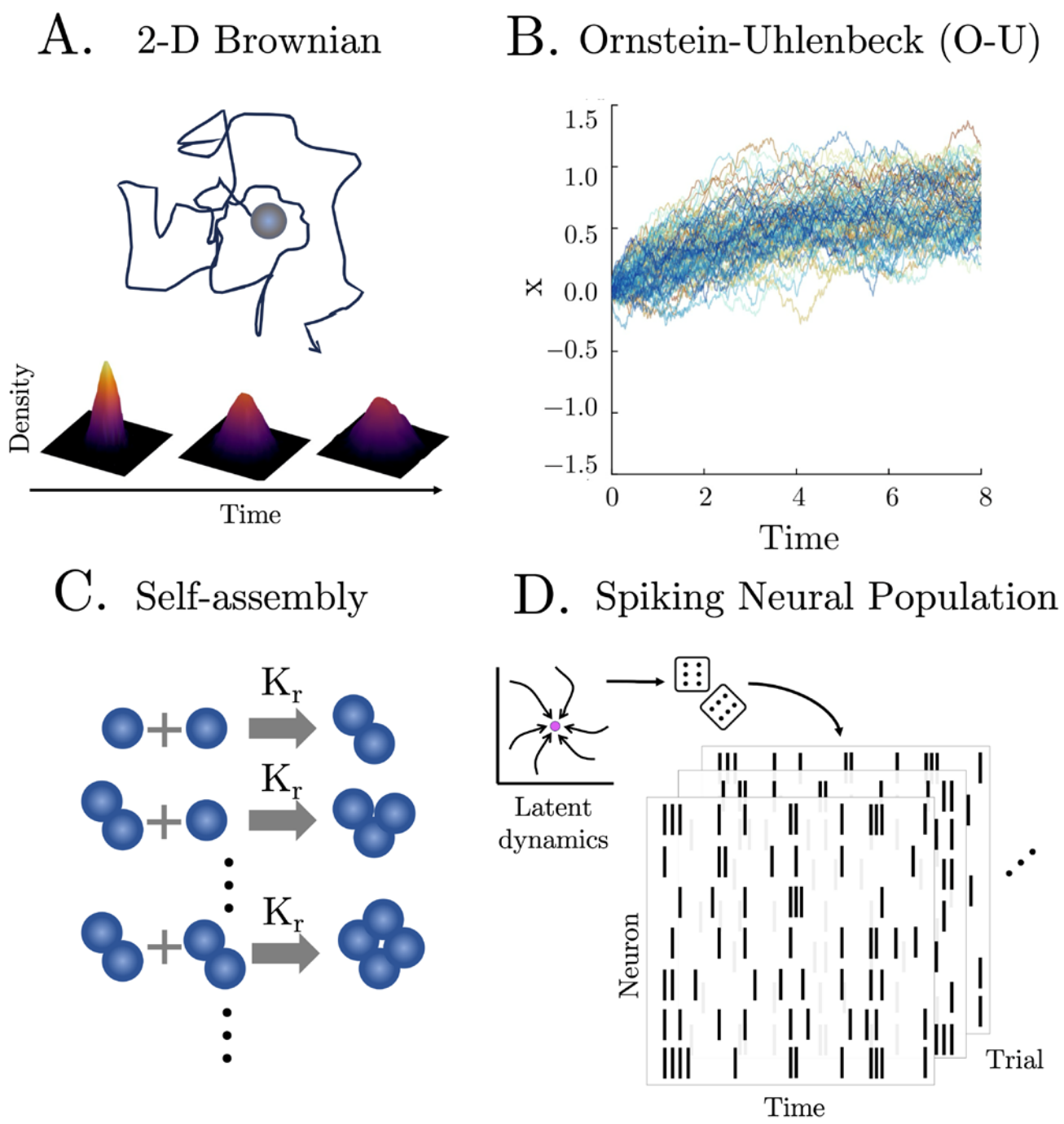

**Figure 3: The diverse systems studied in this work, which include stochastic phenomena (A,B), particle self-assembly (C), and decoding underlying latent dynamics in spiking neural populations (D).**

The third system models particle self-assembly into clusters (Fig. 3C), a ubiquitous phenomenon in engineering, materials science, and biology [51, 52] that has recently been implicated in the anti-tumor efficacy of RNA-based cancer immunotherapies [53, 54]. We simulate 70 particles undergoing Smoluchowski aggregation kinetics [55] and track the evolving cluster-size distribution, with features defined as the number of particles per cluster. This example illustrates that *Edwin* can extract latent dynamics of complex chemical processes directly from nonnegative data, without additional preprocessing.

The final system examines latent dynamics in populations of spiking neurons (Fig. 3D). Recent work in neural computation has shown that low-dimensional population dynamics underlie motor control and decision making [56, 57, 58]. To generate controlled data, we simulate two linear two-dimensional dynamical systems with distinct behaviors: a cyclic attractor resembling motor-cortex dynamics [59, 60, 61] and a stable fixed point associated with decision processes [62, 63, 64, 65]. These latent trajectories are converted into Poisson spiking rates, from which spike trains are sampled. The resulting dataset is tensor-valued, $\mathbf{P} \in \mathbb{R}^{N\times T\times M}$, where $N$ is the number of neurons, $T$ the number of time points, and $M$ the number of experimental conditions corresponding to different initial states. Our goal is to recover a shared low-dimensional dynamical model across these conditions, illustrating that *Edwin* can handle multi-experiment tensor data in addition to standard matrix inputs.

### *Edwin accurately captures low-dimensional latent structure*

First, we gauge the algorithm's ability to reconstruct the high-dimensional input data. To determine the latent space dimension, $K$, we sweep $K \in [1,20]$ and plot the reconstruction loss $\mathcal{L}_{KLD}$ as a function of this parameter without attempting to extract models $\Xi_Z$ and $\Xi_Y$; the latent space dimension we use for subsequent analysis is the smallest $K$ such that $|\log_{10}\mathcal{L}_{KLD}(K+1) - \log_{10}\mathcal{L}_{KLD}(K)| < 0.1$.

For each system, we compare our approach with several reduced-order modeling methods. The first group comprises factorization-based techniques: an exp–log variant of Proper Orthogonal Decomposition (el-POD), Nonnegative Matrix Factorization (NMF), and Dynamic NMF (Dyn-NMF) [66]. POD represents spatiotemporal data as a linear combination of separable spatial and temporal modes, $u(\mathbf{x}, t) \approx \sum_k a_k(t)\phi_k(\mathbf{x})$ [67]. To enable comparison with the exponential DME representation, we apply POD to the logarithm of the data via SVD, truncate to rank $K$, then exponentiate and renormalize. NMF decomposes a nonnegative matrix into low-rank nonnegative factors and has been widely used across scientific domains [68]. Dyn-NMF extends this framework by additionally learning a linear dynamical operator that autoregressively evolves the temporal coefficients; details are provided in Appendix 11.

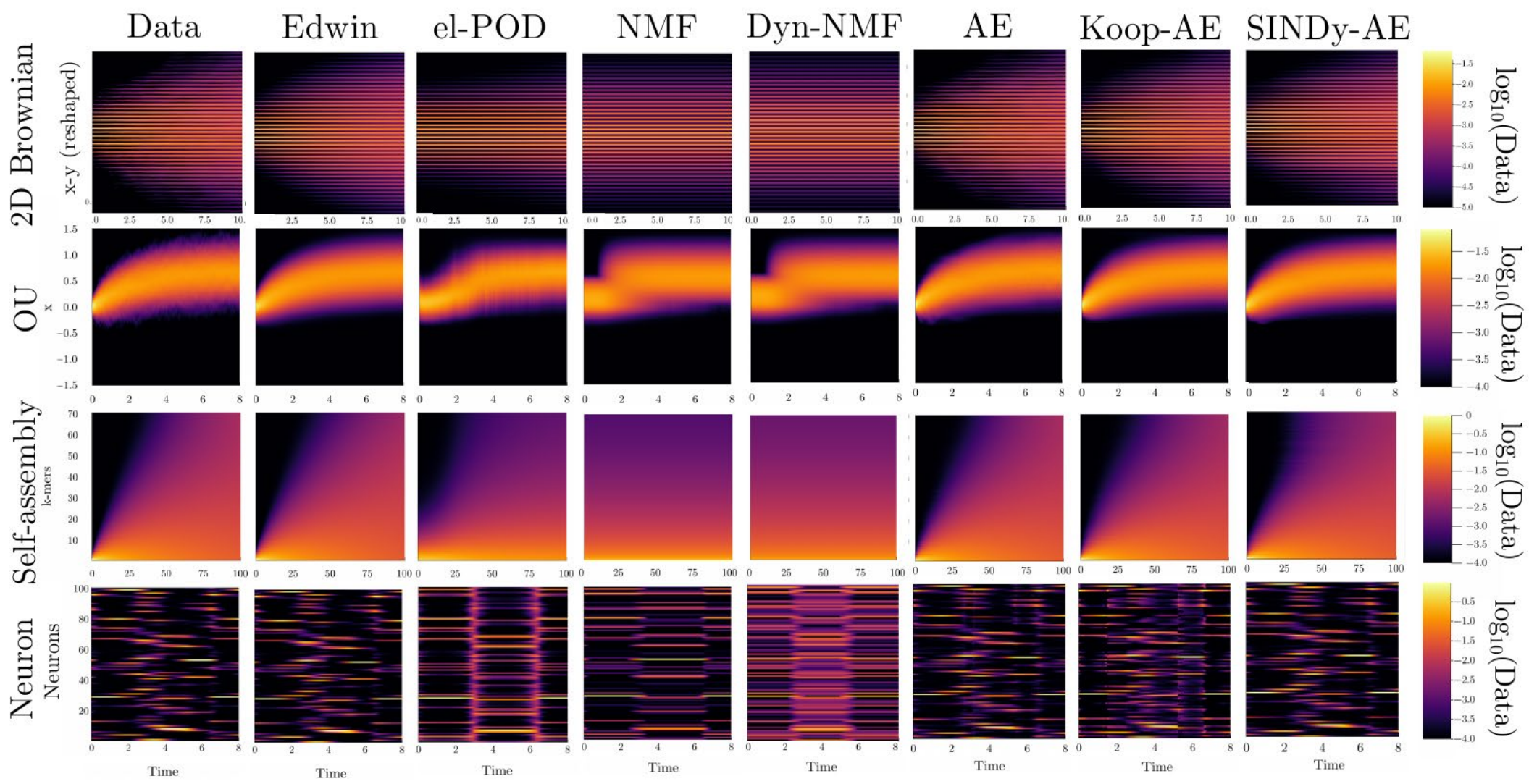

**Figure 4: Reconstruction accuracy across seven methods at the latent dimension $K$ selected by *Edwin*. Methods compared are *Edwin*, exp–log Proper Orthogonal Decomposition (el-POD), Nonnegative Matrix Factorization (NMF), Dynamic NMF (Dyn-NMF), a standard autoencoder (AE), a Koopman autoencoder (Koop-AE), and a SINDy autoencoder (SINDy-AE). For the 2D Brownian system, each $41 \times 41$ spatial grid is reshaped into a column vector for visualization. All methods are evaluated at the same latent dimension chosen by *Edwin* ($K = 1$ for the 2D Brownian and self-assembly systems; $K = 2$ for the Ornstein–Uhlenbeck and neural population systems), enabling direct comparison of reconstruction quality at matched model capacity.**

We also benchmark against several autoencoder-based methods (Appendix 12). A standard autoencoder (AE) compresses the input into a $K$-dimensional latent space and reconstructs it by minimizing the $L_2$ error. We further consider the SINDy-autoencoder (SINDy-AE) [22, 20], which augments the AE with a sparse dynamical model for the latent variables ($\dot{\mathbf{Z}} = \Theta(\mathbf{Z})\Xi$) but does not model feature dependence, limiting interpretability. Finally, Koopman autoencoders (Koop-AE) [69] learn latents $\mathbf{Z}(t)$ together with a linear evolution operator $\mathbf{K}$ satisfying $\mathbf{Z}(t + \Delta t) = \mathbf{K}\mathbf{Z}(t)$. All autoencoder architectures use fully connected encoders and decoders with tanh activations and a softmax output layer to ensure normalization. Networks were trained for 35,000 epochs using the Adam optimizer with learning rate $10^{-3}$, without hyperparameter tuning, as our goal is comparison of reconstruction capability rather than optimization of network performance. Hyperparameters for SINDy-AE and Koop-AE follow their original publications and are summarized in Appendix 12.

At matched latent dimension $K$, *Edwin* and the autoencoder-based methods (AE, SINDy-AE, Koop-AE) reconstruct the data accurately across all systems, whereas the factorization-based methods (el-POD, NMF, Dyn-NMF) perform substantially worse (Fig. 4). The strong performance of neural-network approaches is expected given their universal approximation capability [70]. However, these methods rely on amortized inference via parametric encoders to identify the latent space, which introduces a large number of trainable parameters. In contrast, *Edwin* infers latent variables directly through gradient-based optimization without learning an explicit encoder, achieving comparable reconstruction with substantially fewer parameters. Thus, reconstruction accuracy alone does not distinguish the methods; the principal advantages of *Edwin* lie in its interpretable and consistently recovered latent structure, and in the sparse symbolic models it discovers, as demonstrated in Secs. 4.3 and 4.5.

To further probe these differences, we examine $\mathcal{L}_{KLD}$ as a function of $K$ for all methods (Fig. 5). *Edwin* and all AE-based approaches exhibit saturation of $\mathcal{L}_{KLD}$ beyond the intrinsic latent dimension, converging to a common noise floor set by finite-sample noise and the stochasticity inherent in the kernel density estimates of $p_{it}$. This floor reflects the irreducible discrepancy between the observed noisy distributions and the true underlying dynamics, and cannot be reduced by adding further latent dimensions. In contrast, the factorization-based methods (el-POD, NMF, Dyn-NMF) continue to decrease below this floor at sufficiently large $K$. Rather than reflecting superior compression, this behavior indicates overfitting to the specific noise realization in $p_{it}$: their linear or nonnegative-linear decoders, lacking the inductive bias of an exponential-family or neural-network model, use additional dimensions to fit noise rather than to capture distributional structure. This artificial inflation of the latent dimension has important implications for both interpretability and forecasting, as high dimensionality makes both efforts more challenging [71, 72].

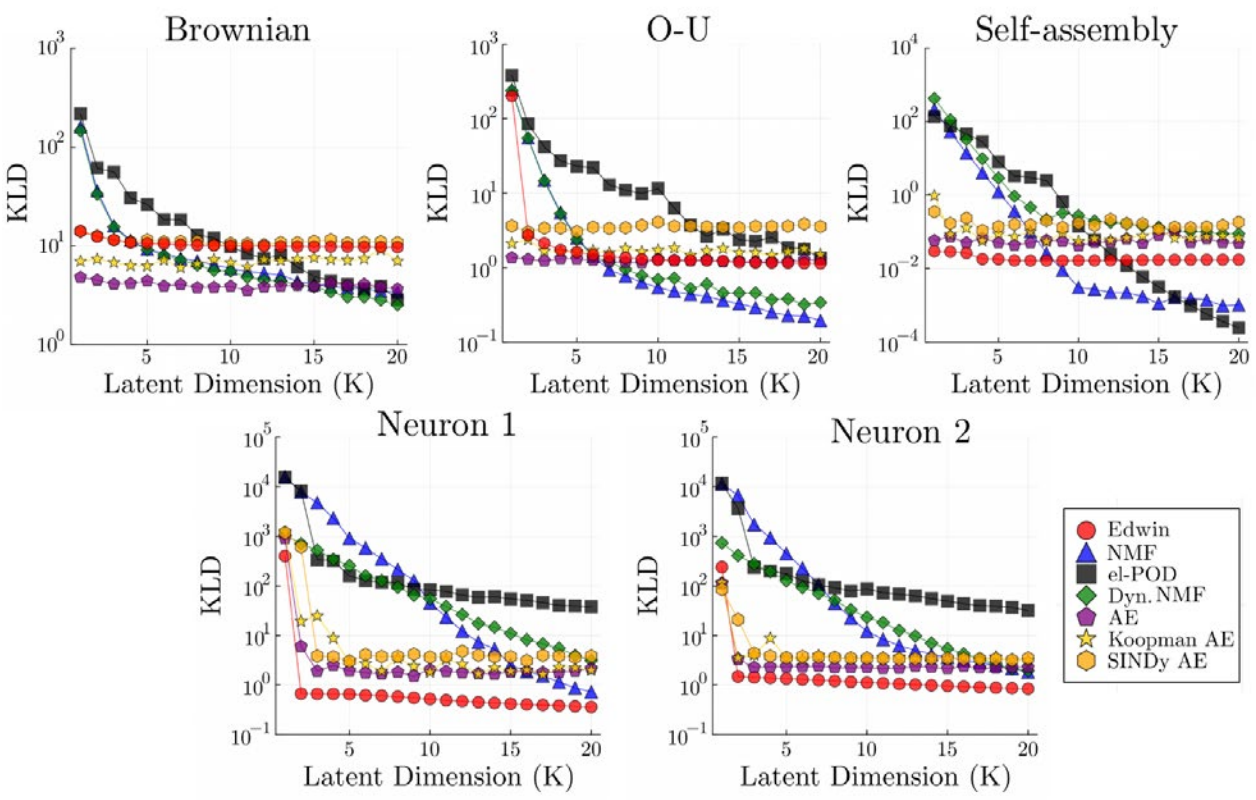

**Figure 5: Reconstruction loss $\mathcal{L}_{KLD}$ as a function of latent dimension $K$ for all methods. Factorization-based methods (el-POD, NMF, Dyn-NMF) require substantially larger $K$ to approach the same reconstruction quality, and ultimately fall below the noise floor at high $K$, indicating overfitting to noise in $p_{it}$ rather than genuine compression. AE-based methods (AE, Koop-AE, SINDy-AE) saturate at the same noise floor as *Edwin* at comparable $K$, consistent with their nonlinear approximation power; however, unlike DME, none of these methods jointly learn feature-dependent latents Y and an interpretable sparse model $\Xi_Y$ that couples those latents to external metadata, limiting the physical interpretability of the learned representations.**

### *Sparse model discovery of latent space variables*

Next, we demonstrate that *Edwin* extracts interpretable governing models for the latent variables. This capability contrasts with many compression methods that prioritize reconstruction accuracy while leaving the latent dynamics unspecified. Methods that do model latent evolution frequently parameterize it using neural networks [21, 73], which limits interpretability. By jointly learning the feature-dependent model $\mathbf{Y} = \Theta_Y(\mathbf{x})\Xi_Y$, *Edwin* assigns direct physical meaning to each latent coordinate, as illustrated below. At the end of this section, we compare directly against SINDy-AE, the most relevant neural-network baseline (Fig. 6).

*2D Brownian*. For the Brownian system (Fig. 6A), *Edwin* correctly identifies a Gaussian distribution with zero mean, with feature model $Y(\mathbf{x}) = Y(x_1, x_2) \approx 3.21x_1^2 + 3.35x_2^2$. The inferred dynamics is $\dot{Z} \propto -Z^2$, whose solution $Z(t) \propto 1/t$ implies that the variance grows linearly in time. Together, these results recover the known diffusive behavior directly from data. Notably, despite low-resolution and noisy inputs, the inferred coefficients indicate nearly isotropic diffusion, as the nonzero entries of $\Xi_Y$ are approximately equal.

*OU process*. The latent space models for the OU system (Fig.6B) is richer, as it is two-dimensional. In principle, the dynamics of such a system could exhibit a multitude of behaviors, ranging from fixed points to limit cycles [74]. *Edwin* uncovers that the distribution evolves as a Gaussian, with $Y_1(x) \propto x$ and $Y_2(x) \propto x^2$. Thus, the dynamics dictates how the mean and standard deviation evolve with time. From a second-order library the equations are:

$$\begin{aligned} \dot{Z}_1 &= 0.39Z_1 + 0.85Z_2 - 0.44Z_1Z_2 \\ \dot{Z}_2 &= -0.39Z_2 - 0.29Z_2^2 \end{aligned} \tag{14,15}$$

From the equations, we can see that there is a stable fixed point, which, taken with the feature model, indicates that the distribution settles to a final mean and variance.

*Self-assembling particles*. For the self-assembly dynamics (Fig.6C), a system of 70 coupled nonlinear ODEs, *Edwin* reduces the dynamics to a single ODE of the diffusive type: $\dot{Z} \propto -Z^2$. The relevant features in this system are the cluster sizes (i.e., how many particles are per cluster), which we denote by $k$; the corresponding feature-dependent latent model is $Y(k) \propto k$. Thus, if $N(t)$ denotes the total number of clusters at a time $t$, then the effective distribution for clusters of size $k$ at time $t$ is $n_k(t)/N(t) \propto e^{-Z(t)k}$. From the standpoint of dynamic maximum entropy, this immediately suggests that the total number of particles ($M = \sum_k k\, n_k$) and the total number of clusters ($N = \sum_k n_k$) are the most relevant quantities for modeling the evolving distribution; by scaling $N$ and $M$ by $1/N$ and letting $p_k = n_k/N$, our model extracted from data can be interpreted as maximizing entropy subject to constraining the normalization condition ($\sum_k p_k$) and the first moment $\sum_k k\, p_k$. *Edwin* was able to extract this information without any prior assumptions. Such a drastic reduction in dimensionality will likely be useful for future modeling and simulation efforts for these complex systems, and indeed motivates the application of our framework to the experimental RNA–liposome aggregation system studied in Section 4.7, where a similar self-assembly process governs cluster growth.

*Neural population*. From a second-order library in Z, *Edwin* correctly identifies that the dynamics of each neural population is linear (Fig. 6D). The predicted latent trajectories closely match the true dynamics after affine alignment (Fig. 6D). This agreement is further quantified by the eigenvalues: the first population, which exhibits oscillatory behavior, yields $\lambda_{true} \approx \pm 0.9487i$ versus $\lambda_{pred} \approx \pm 0.9488i$; the second population, with apparent fixed-point behavior, yields $\lambda_{true} \approx -0.1576 \pm 0.7048i$ versus $\lambda_{pred} \approx -0.1577 \pm 0.7026i$. Importantly, for the second population, the true latent dynamics had not converged to the fixed point before encoding into firing rates; *Edwin* infers that a fixed point exists without it being present in the training data.

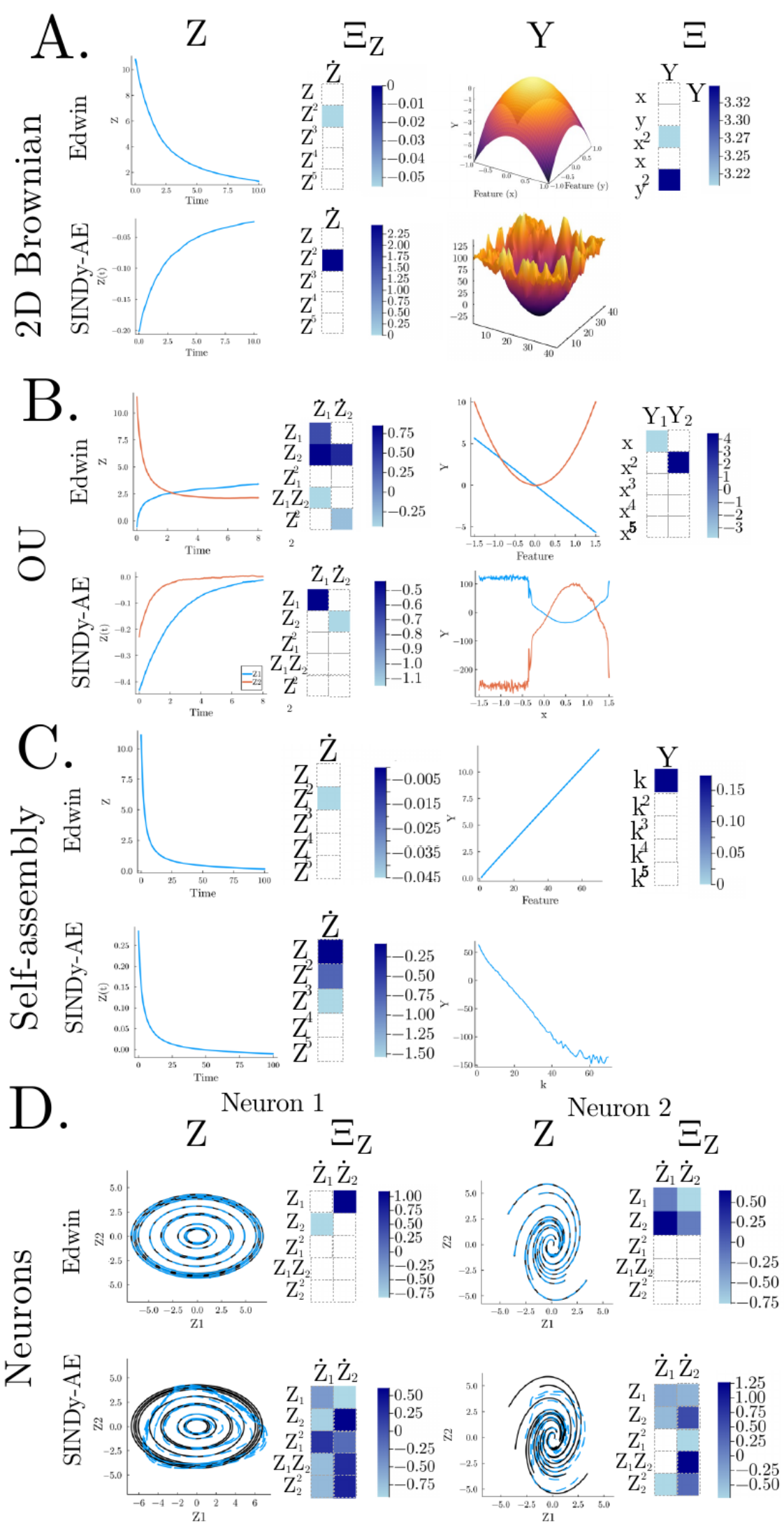

**Figure 6: Latent variables and inferred models for *Edwin* (top row of each panel) and SINDy-AE (bottom row), across all systems. For *Edwin*, the time-dependent latents Z, feature-dependent latents Y, and sparse coefficient matrices ($\Xi_Z, \Xi_Y$) are shown; for SINDy-AE, the time-dependent latents Z and sparse dynamics model $\Xi_Z$ are shown, together with the feature-space representation W obtained by linear regression of the decoder output onto Z (see text). In the model visualizations, rows correspond to candidate terms and columns to latent variables; blank entries enclosed by dashed boxes indicate coefficients pruned to zero by SLIC. In panel (D), latent trajectories (blue) are shown alongside the true dynamics (black), after finding the affine alignment $\mathbf{Z}_{true} \approx \mathbf{A}\mathbf{Z}_{pred} + \mathbf{t}$. The SINDy-AE models shown are the pre-adaptation (training-only) weights used as the starting point for the forecasting comparison in Sec. 4.5.**

Figure 6 also shows the latents and models recovered by SINDy-AE under identical conditions. These are the pre-adaptation, training-only models used as the starting point for the forecasting comparison in Sec. 4.5. Several differences are apparent. First, SINDy-AE does not learn a feature-dependent model $\mathbf{Y}$; to enable a comparison, we extract a linear approximation $\mathbf{W}$ of the decoder by regressing the decoder output onto $\mathbf{Z}$ via ordinary least squares, so that $\psi(\mathbf{Z}) \approx \mathbf{W}\mathbf{Z}$. The columns of $\mathbf{W}$ are substantially less smooth than the feature-dependent latents $\mathbf{Y}$ recovered by *Edwin*: for the Brownian system, *Edwin* produces a smooth paraboloid $Y(\mathbf{x}) \propto x_1^2 + x_2^2$ consistent with a Gaussian distribution, whereas the corresponding SINDy-AE decoder surface is noticeably more irregular. A similar contrast is observed for the self-assembly system, where *Edwin* recovers a smooth linear $Y(k) \propto k$ while the SINDy-AE decoder is non-monotone and noisy. This difference reflects the fact that the SINDy-AE decoder is a nonlinear neural network optimized for reconstruction accuracy, not for producing *de novo* structured feature representations.

Second, the scale and sign of the SINDy-AE latents are arbitrary: the latent space is defined only up to invertible transformations of the encoder-decoder pair, so the recovered $\mathbf{Z}$ carries no inherent physical meaning without additional post-hoc analysis [47]. The feature-dependent model $\mathbf{Y} = \Theta_Y(\mathbf{x})\Xi_Y$ in *Edwin* resolves this by anchoring each latent coordinate to external metadata $\mathbf{x}$, providing unambiguous physical interpretation without post-hoc interrogation.

Third, for the neural population systems, *Edwin* recovers sparser $\Xi_Z$ coefficients than SINDy-AE, reflecting a more parsimonious model of the underlying linear dynamics. We further quantify the differing geometric structure of the recovered latent spaces using the trajectory tanglement metric [60, 59], which measures the similarity of flow fields as neighboring trajectories approach one another; a lower tanglement score indicates smoother flow fields. For a pair of trajectories $(i, j)$, velocities $\dot{\mathbf{Z}}$ are computed from $\Theta(\mathbf{Z})\Xi_Z$, and tanglement is defined as

$$Q_{ij} = \max_{t,t'} \frac{\|\dot{\mathbf{Z}}_i(t) - \dot{\mathbf{Z}}_j(t')\|_2^2}{\|\mathbf{Z}_i(t) - \mathbf{Z}_j(t')\|_2^2 + \varepsilon}, \quad (16)$$

with $\varepsilon = 10^{-6}$, and reported as the maximum across all trajectory pairs. Across multiple realizations ($n = 5, 10$, for SINDy-AE and *Edwin*, respectively), *Edwin* yields median tanglement values of 1.08 (interquartile range (IQR) ≈ 0.08) (Neuron 1) and 0.63 (IQR ≈ 0.04) (Neuron 2), compared to 4.58 (IQR ≈ 1.62) and 2.21 (IQR ≈ 0.38) for SINDy-AE, indicating that SINDy-AE's latent space is considerably more tangled. Importantly, *Edwin*'s tanglement is also lower for all pairs of trajectories (Fig. S3).

### *Robustness to initialization and latent dimensionality*

We assessed the robustness of the proposed framework with respect to (i) random initialization of latent variables and (ii) mild over-specification of the latent dimensionality. Results are summarized in Figs. 7 and 8.

Because the latent trajectories are optimized jointly with the governing equations, different initializations could in principle lead to distinct local minima. To evaluate this sensitivity, we performed ten independent runs for each system using time- and feature-dependent latent variables initialized from a standard normal distribution. For each set of runs, we report the median model (Fig. 7). Across all systems considered, the inferred governing equations and qualitative latent dynamics are highly consistent across initializations, indicating that the optimization landscape contains a dominant basin corresponding to the physically relevant solution. This suggests that the recovered models are not artifacts of a particular initialization.

We further examined the effect of modestly overestimating the latent dimension by repeating the analysis with $K + 1$ latent coordinates (Fig. 8). While under-specifying $K$ leads to a clear degradation in reconstruction quality (Fig. 5), our primary interest here is whether mild over-specification manifests as redundancy in the recovered latents and models, or instead produces qualitatively different dynamics. As we show below, mild over-specification primarily introduces redundant coordinates rather than qualitatively new dynamics.

For the Brownian and self-assembly systems, the additional coordinate becomes strongly correlated with existing ones, effectively embedding the original low-dimensional dynamics in a higher-dimensional space. In the neuronal system, the three-dimensional latent trajectories clearly exhibit attractor structure similar to that obtained at the optimal $K$, although the resulting models are somewhat less sparse. For the Ornstein–Uhlenbeck system, the extra coordinate ($Z_2$, up to arbitrary labeling) is driven toward an approximately constant value with a weak coupling to the primary latent, indicating that it carries little dynamical information.

In this latter case, fully eliminating the redundant coordinate would render the regression library nearly rank-deficient, since columns of the feature matrix $\Theta_Z$ become highly collinear. The regression therefore retains a small stabilizing coefficient rather than removing the term entirely. Overall, these results indicate that modest overestimation of $K$ produces degenerate embeddings of the true dynamics rather than fundamentally different models.

Taken together, Figs. 7 and 8 demonstrate that the inferred models are stable with respect to both initialization and mild mis-specification of the latent dimension. In contrast, SINDy-AE exhibits substantially higher run-to-run variability: across multiple random initializations, the recovered latent trajectories and sparse models differ qualitatively between runs (Fig. S1). This discrepancy is consequential: whereas *Edwin*'s consistent convergence reflects a dominant optimization basin tied to the physical structure of the data, SINDy-AE's sensitivity to initialization means the recovered model may be an artifact of a particular random seed rather than a reflection of the underlying system dynamics, complicating physical interpretation and undermining confidence in the discovered governing equations.

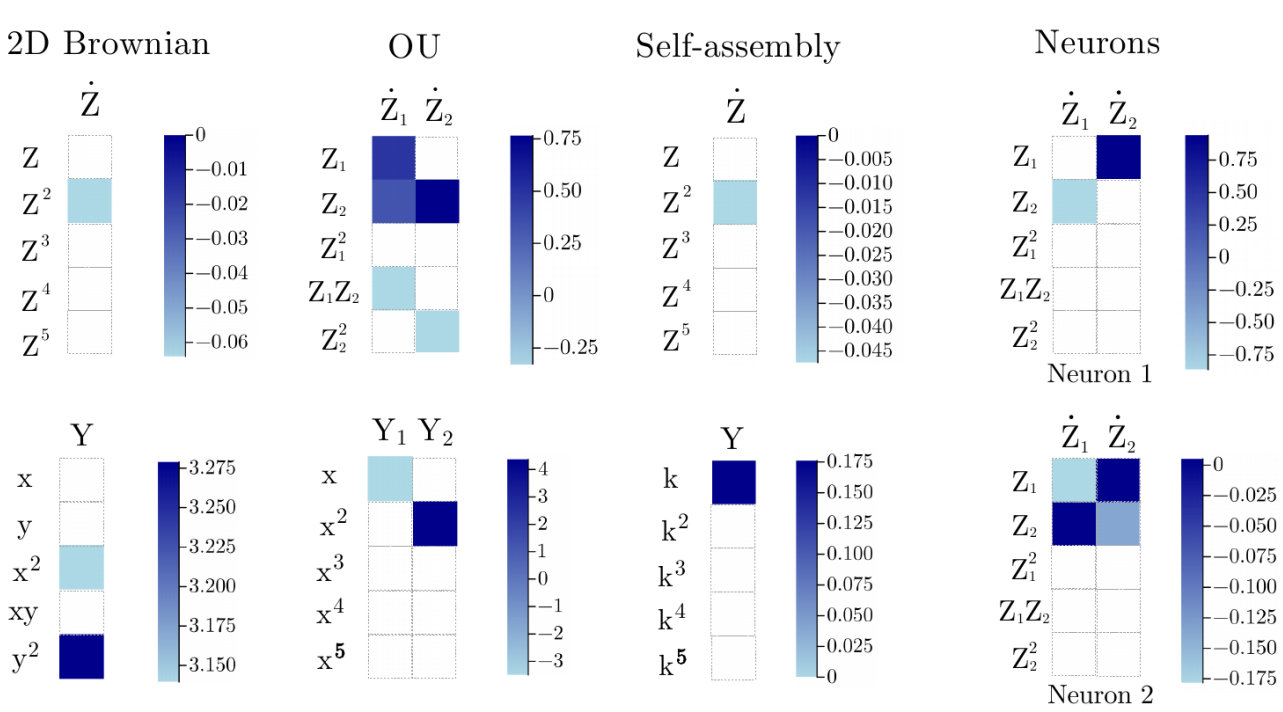

**Figure 7: Robustness of inferred latent models to random initialization. For each system, ten independent runs were performed with time- and feature-dependent latent variables initialized from a standard normal distribution. The model shown is the median across these runs. The recovered governing equations and qualitative latent dynamics are highly consistent across all runs, indicating that the method is robust to initialization.**

### *Discovered dynamics generalize to unseen conditions*

A central goal of model discovery is to predict system evolution beyond the data used for training. We distinguish between two related but conceptually distinct tasks. *Forecasting* refers to predicting future states under the same dynamical law given partial observations of a trajectory. *Adaptation*, by contrast, refers to predicting behavior under altered conditions, such as changes in parameters, initial states, or boundary conditions, where the underlying dynamics may differ from those seen during training. Successful adaptation requires that the learned model capture transferable structure rather than merely reproducing previously observed trajectories.

To assess both capabilities, we evaluate the discovered models on the same systems studied above under a range of previously unseen conditions. These perturbations induce distribution shifts relative to the training data and therefore test whether the inferred dynamics generalize to new regimes rather than interpolate within the original dataset.

Specifically, prefixes of length $\{1,50,100,150\}$ time points are provided to the model, where a prefix of length 1 corresponds to no update. For each prefix, only the components governing latent evolution are updated before predicting the remainder of the trajectory.

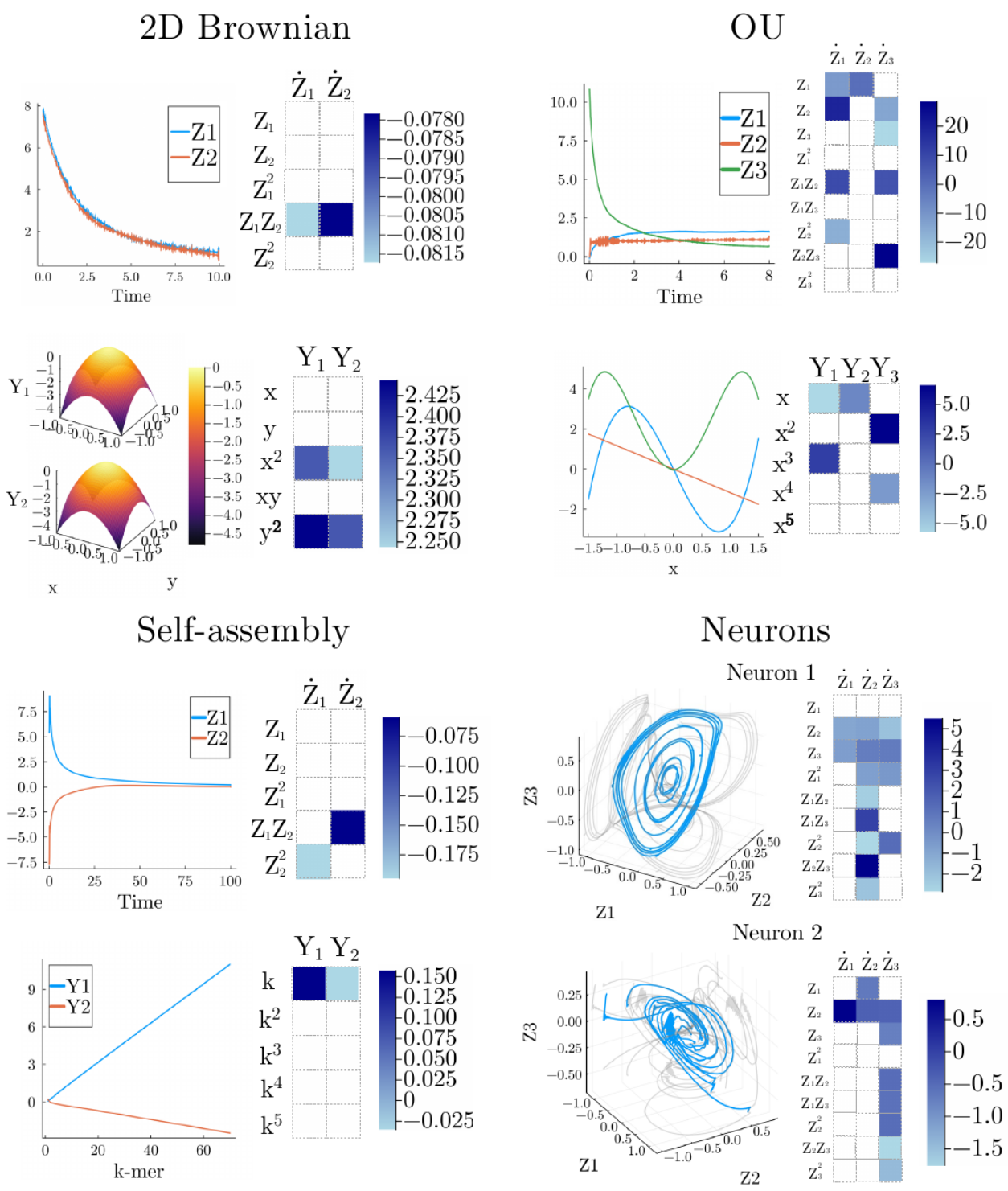

**Figure 8: Effect of mild over-specification of latent dimensionality ($K + 1$). Across systems, the additional coordinate primarily introduces redundant or weakly active directions rather than qualitatively new dynamics. For the Brownian and self-assembly systems, the extra latent becomes strongly correlated with existing coordinates. In the neuronal system, the three-dimensional trajectories clearly exhibit attractor structure similar to the optimal-$K$ case (light gray curves show projections onto orthogonal planes). For the Ornstein–Uhlenbeck system, the additional coordinate ($Z_2$, up to arbitrary labeling) is nearly constant with weak coupling to the primary latent, indicating limited dynamical relevance. These results suggest that modest overestimation of $K$ yields degenerate embeddings of the underlying dynamics rather than fundamentally different models.**

For *Edwin*, the nonzero coefficients of $\Xi_Z$ and the latent trajectories $\mathbf{Z}$ are updated jointly using the available prefix data, while $\mathbf{Y}$ and $\Xi_Y$ are held fixed. For autoencoder-based baselines, a hyperparameter grid search is performed before adaptation, after which only the latent evolution operators ($\Xi_Z$ for SINDy-AE, the Koopman matrix $\mathbf{K}$ for Koopman-AE) are updated with encoder and decoder weights held fixed. Full details of the adaptation procedure, including epoch counts and hyperparameter grids for each architecture, are given in Appendix 12.

Adaptation accuracy is quantified using the maximum relative $L_1$ error:

$$\max_t \frac{\|\mathbf{P}_{true}(t)-\mathbf{P}_{pred}(t)\|_1}{\|\mathbf{P}_{true}(t)\|_1} \tag{17}$$

where $\mathbf{P}_{true}$ and $\mathbf{P}_{pred}$ denote the true and predicted data, respectively. This metric emphasizes worst-case deviation over the forecast horizon.

Figure 9 shows representative forecasts and the maximum relative error across methods. Zero-shot forecasting performs poorly for most approaches, including our own. This behavior is expected because the underlying dynamics in the test data differ from those seen during training, so a model with fixed parameters cannot accurately predict the new trajectory. As the prefix length increases, forecasting accuracy improves for *Edwin*, indicating that the discovered dynamical structure can be rapidly adapted using limited new data.

In contrast, the maximum relative forecasting error decreases only marginally for Dyn-NMF, SINDy-AE, and Koopman-AE. This limited improvement may reflect inaccuracies in the learned dynamical models under distribution shift. More generally, for neural-network-based approaches, it may also reflect the data-hungry nature of these models, since only a small prefix of a single trajectory is available for adaptation.

Across nearly all systems and prefix lengths, *Edwin* achieves substantially lower forecasting errors than competing methods, typically by at least an order of magnitude in maximum relative $L_1$ error. Notably, for the Brownian, Ornstein–Uhlenbeck, and self-assembly systems, training was performed using only a single trajectory. Successful forecasting under new conditions therefore indicates that *Edwin* infers transferable governing dynamics rather than memorizing specific observations. Together, these results demonstrate that *Edwin* not only yields interpretable models but also enables accurate prediction of system evolution under previously unseen conditions with minimal additional data.

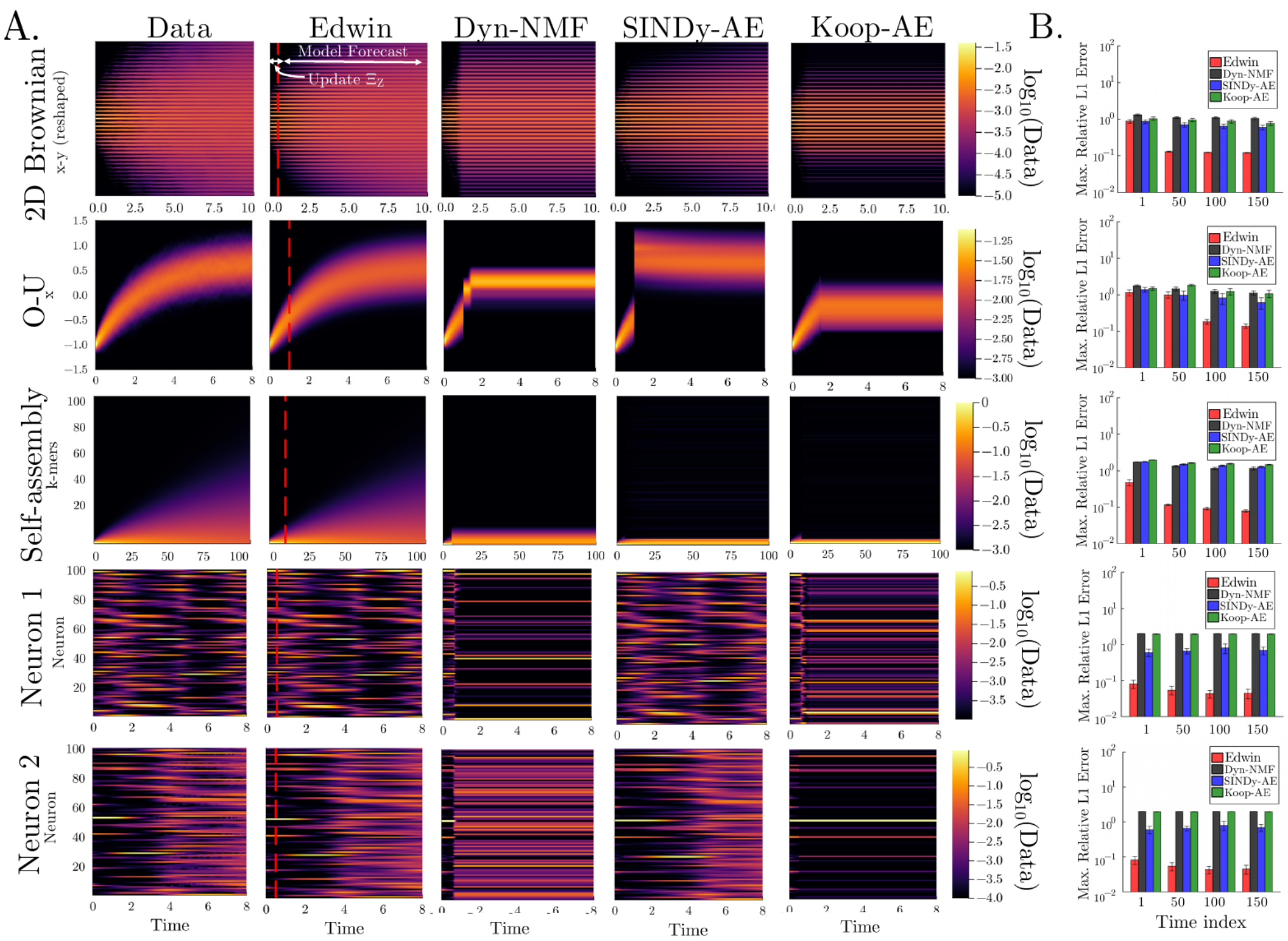

**Figure 9: Generalization of discovered dynamics to unseen conditions via limited adaptation. (A) Example predictions for each system. The available prefix data (left of the red dashed line) is used to jointly update Z and the nonzero coefficients of $\Xi_Z$ while Y and $\Xi_Y$ are held fixed; the remainder of the trajectory is then predicted using the updated model. (B) Maximum relative $L_1$ adaptation error as a function of prefix length. Data are displayed as mean ± s.t.e. (n=10).**

*Extracting effective dynamics from low-rank random neural networks*

To determine whether the framework recovers intrinsic dynamical structure rather than fitting high-dimensional activity, we analyzed recurrent networks of the form

$$\dot{x}_i = -x_i + \sum_{j=1}^{N} W_{ij}\ \varphi(x_j), \tag{18}$$

where $x_i(t)$ denotes the activity of neuron $i$, $\varphi(\cdot) = \mathrm{ReLU}(\cdot)$ is the nonlinear activation function, and $W$ is the recurrent connectivity matrix [76]. We use networks of $N = 100$ neurons throughout.

We constructed $W$ as the sum of a low-rank structured component and full-rank random perturbations,

$$W = \frac{1}{N} U U^\top + \frac{\varepsilon}{\sqrt{N}} G, \tag{19}$$

where $U \in \mathbb{R}^{N\times r}$ defines an $r$-dimensional structured subspace, $G_{ij} \sim \mathcal{N}(0,1)$ are independent Gaussian entries, and $\varepsilon$ controls the strength of the random component. The normalization factors ensure a well-defined large-$N$ limit.

In the absence of random perturbations ($\varepsilon = 0$), the network dynamics are confined to the $r$-dimensional subspace spanned by the columns of $U$. Specifically, defining the collective coordinates

$$\kappa(t) = \frac{1}{N} U^\top \varphi(x(t)) \in \mathbb{R}^r, \tag{20}$$

in the large-$N$ limit, the network activity can be approximated as [76]

$$x(t) \approx U\kappa(t), \tag{21}$$

and the high-dimensional dynamics reduce to a closed system for $\kappa(t)$. This is an approximate mean-field expression: it holds exactly only as $N \to \infty$ and $\varepsilon \to 0$, with corrections of order $1/\sqrt{N}$ for finite networks. For small but finite $\varepsilon$, this low-dimensional structure persists as an attracting subspace, with the random component introducing fluctuations transverse to the structured directions. As $\varepsilon$ approaches unity, however, the random component is expected to dominate and the low-rank structure to be progressively destroyed [76], a prediction we can directly test. This construction therefore provides a controlled setting in which the intrinsic dynamical dimension $r$ is known *a priori*.

To analyze the resulting trajectories within the probabilistic representation used throughout the manuscript, neural population activity was expressed in normalized form at each time point. The learned latent dynamics are then propagated forward and mapped through a softmax transformation to ensure nonnegativity and unit normalization, allowing direct comparison with the empirical distributions. This step is consistent with the single-neuron probabilistic example introduced earlier and enables application of the DME formalism to recurrent population activity without altering the underlying dynamical structure.

We then swept the candidate latent dimension $K$ and computed the Kullback–Leibler divergence between empirical and reconstructed trajectory distributions (Fig. 10B). We observe that the $K$ at which the KLD begins to saturate grows with the ground-truth rank $r$, suggesting that the framework allocates additional latent capacity in proportion to the intrinsic dimensionality of the dynamics. This relationship persists across low noise levels $\varepsilon \in \{10^{-5}, 10^{-1}\}$. Consistent with the expectation above, as $\varepsilon \to 1$ the low-rank structure is destroyed and no clear saturation is observed across the range of $K$ tested (Fig. 10B), confirming that the framework is sensitive to the presence or absence of low-rank structure.

To assess whether the recovered latent variables align with the collective coordinates defined in Eq. (20), we compared the inferred coordinates $Z(t) \in \mathbb{R}^K$ with $\kappa(t) \in \mathbb{R}^r$, setting $K = r$ throughout. Because both sets of coordinates are defined only up to invertible linear transformations, we do not expect individual components to match directly. We therefore quantified alignment using canonical correlation analysis (CCA), which is invariant to such linear reparameterizations.

Given centered time series matrices $Z \in \mathbb{R}^{K\times T}$ and $\kappa \in \mathbb{R}^{r\times T}$, CCA seeks vectors $a \in \mathbb{R}^K$ and $b \in \mathbb{R}^r$ that maximize the correlation

$$\rho = \max_{a,b} \frac{a^\top \Sigma_{Z\kappa} b}{\sqrt{a^\top \Sigma_{ZZ} a}\sqrt{b^\top \Sigma_{\kappa\kappa} b}} \tag{22}$$

where $\Sigma_{ZZ}$, $\Sigma_{\kappa\kappa}$, and $\Sigma_{Z\kappa}$ are covariance and cross-covariance matrices. Solving this problem yields pairs of canonical variables whose correlations $\{\rho_1, \dots, \rho_{\min(r,K)}\}$ quantify the alignment between the two subspaces, with $\rho_i = 1$ indicating perfect correspondence along a shared direction and $\rho_i = 0$ indicating orthogonality.

Across ground-truth ranks $r = 2$–$5$ with $K = r$, representative trajectory comparisons show that, after linear alignment, $Z(t)$ closely tracks $\kappa(t)$ (Fig. 10C). This visual agreement is quantified by CCA: canonical correlations are consistently high across all ranks tested, while the shuffled control, in which the time indices of $Z(t)$ are randomly permuted, yields correlations of approximately 0.2 across all ranks (Fig. 10D), confirming that the alignment reflects shared temporal structure rather than a statistical artifact.

Together, these results indicate that the framework recovers the collective coordinate subspace defined by Eq. (20), rather than fitting high-dimensional activity patterns induced by random connectivity.

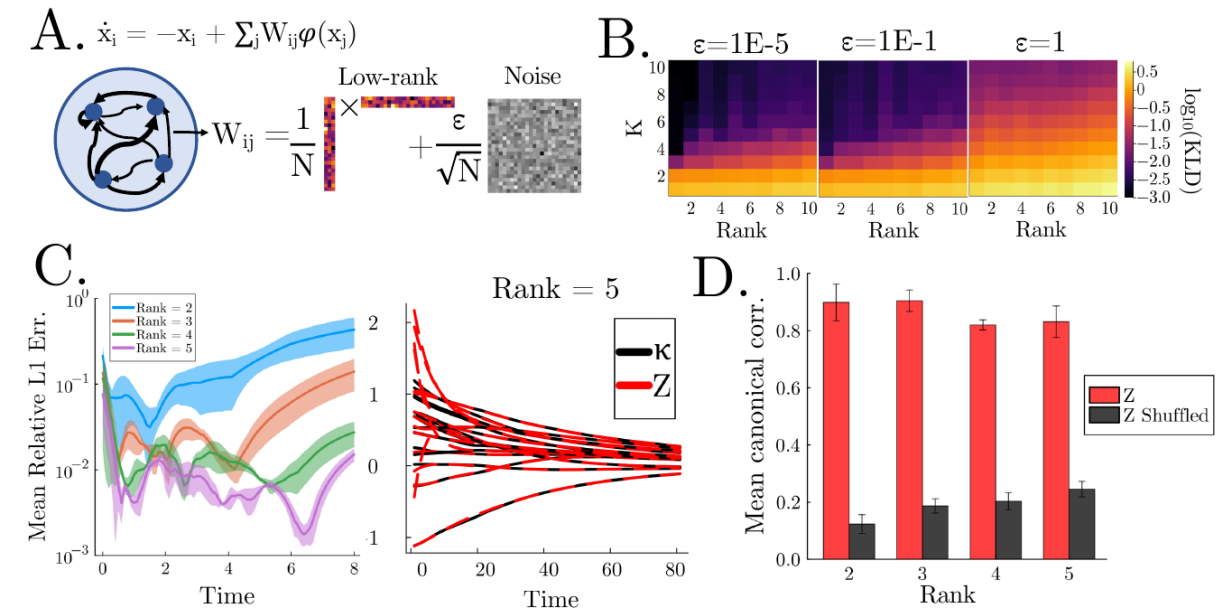

**Figure 10: Recovery of intrinsic low-rank dynamics in recurrent networks. (A) Recurrent dynamics $\dot{x}_i = -x_i + \sum_j W_{ij}\,\varphi(x_j)$ with $W = \frac{1}{N} U U^\top + \frac{\varepsilon}{\sqrt{N}} G$, where $U \in \mathbb{R}^{N\times r}$ defines an $r$-dimensional structured subspace and $G_{ij} \sim \mathcal{N}(0,1)$ adds full-rank noise. The intrinsic coordinates are $\kappa(t) = \frac{1}{N} U^\top \varphi(x(t))$. (B) $\log_{10}$ KLD as a function of candidate latent dimension $K$ (x-axis) for ground-truth ranks $r = 1$–$5$ (separate curves) and noise levels $\varepsilon \in \{10^{-5}, 10^{-1}, 1\}$ (separate panels). The saturation point of the KLD grows with $r$ at low noise, consistent with the framework identifying the intrinsic rank. At $\varepsilon = 1$, the low-rank structure is destroyed and no clear saturation is observed, as expected from [76]. (C) Mean relative $L_1$ error between $\kappa(t)$ and a linearly aligned latent representation $Z(t)$ for $K = r$. Example shown for $K = r = 5$. (D) Mean canonical correlations between $Z(t)$ and $\kappa(t)$ across $K = r = 2$–$5$; values near 1 indicate perfect subspace alignment (shuffled control in black).**

### *Discovering dynamics from fluorescence images of aggregating RNA–liposome complexes*

To evaluate performance in a noisy experimental setting with large observable support and unknown governing equations, we apply *Edwin* to fluorescence microscopy time series of aggregating RNA–liposome complexes [53, 54]. Unlike the simulated self-assembly systems considered previously, this dataset contains experimental measurement noise, finite sampling effects, and no known ground-truth dynamical model.

Clusters were imaged in real time by labeling the liposome component with fluorescent lipid-affinity dyes and tracking aggregate growth over 28 time points (one per minute). Fluorescence microscopy naturally produces intensity measurements proportional to the number of labeled particles, which can be directly converted into relative abundances of cluster sizes (k-mers) without absolute calibration. This relative representation is nonnegative and normalized at each time point, making it directly suitable as input to DME, and is robust to experiment-to-experiment fluctuations in imaging conditions. This yields empirical distributions over $k \in \{1, \dots, 12\}$ at each time point, forming a dataset $\mathbf{P} \in \mathbb{R}^{12\times 28}$; we retain cluster sizes up to $k_{\max} = 12$, beyond which the mean relative abundance falls below $5 \times 10^{-3}$ across all time points.

This setting poses several challenges: large observable support ($k_{\max} = 12$), limited temporal samples ($T = 28$),

experimental noise in fluorescence quantification, and unknown aggregation kinetics.

We first estimate the latent dimension by sweeping $K \in [1,10]$ without extracting dynamical models and computing the Kullback–Leibler divergence (KLD). The KLD decreases by less than 3% (on a log scale) across this range (Fig. 11A), indicating that despite the large observable space, the data lie close to a low-dimensional dynamical manifold. We therefore select $K = 1$, allowing the method to determine intrinsic complexity adaptively rather than fixing it by assumption. Notably, large observable support does not imply high latent dimension; instead, the evolution of the full 12-dimensional distribution is effectively governed by a single coordinate.

With $K = 1$, we extract feature-dependent latents of the form $Y(k) = \Theta_Y(k)\Xi_Y$. The candidate library includes monomials up to fifth order $(k, k^2, \dots, k^5)$ and logarithmic terms $(\log k, k\log k, k^2\log k, k^3\log k)$, enabling identification of a broad class of exponential-family forms including Poisson-, Gamma-, Gaussian-, exponential-, and power-law–type distributions.

From this nontrivial hypothesis space, the algorithm selects $\log(k)$ as the dominant feature-dependent term (Fig. 11B). This implies that the evolving distribution takes the approximate form:

$$P(k,t) \propto k^{-Z(t)} \tag{23}$$

i.e., a power-law distribution with a time-dependent scaling exponent.

Using a third-order polynomial library for the latent dynamics, SLIC identifies a logistic-type model:

$$\dot{Z} = rZ\left(1 - \frac{Z}{\tau}\right) \tag{24}$$

where $r$ is a growth rate and $\tau$ is the steady-state exponent. This indicates that the scaling exponent evolves nonlinearly and relaxes toward a stable fixed point.

Despite the limited number of time points and experimental noise, the resulting model reproduces both the full distribution and the first three moments of the data (Fig. 11C–D), even though moment constraints were not imposed during training. This demonstrates that the discovered structure captures global statistical properties rather than merely fitting local features.

Power-law scaling is well known in clustering phenomena [55, 77] and is associated with self-similarity and homogeneity of the cluster–cluster interaction kernel in Smoluchowski-type coagulation models (Eq. (25)). In particular, homogeneity of the kernel of the form $K_{\lambda i,\lambda j} = \lambda^{\nu} K_{i,j}$ leads to scale-free cluster-size distributions with evolving power-law exponents [55]. The value of the scaling exponent $Z(t)$ and its steady-state $\tau$ can therefore be interpreted in terms of the dominant aggregation mechanism; in particular, the scaling exponent has been linked to distinguishing diffusion-limited from reaction-limited regimes [78, 55].

Here, we observe $Z(0) \approx 2.75$, consistent with a diffusion-limited aggregation regime, and $\tau \approx 1.41$, characteristic of a reaction-limited steady state. This suggests that the dominant clustering mechanism shifts over time, potentially reflecting structural rearrangements in the mRNA–liposome complexes as aggregation proceeds [78]. Importantly, this mechanistic insight emerges in low-dimensional coordinates without imposing a coagulation model *a priori*.

We next examine how the time-dependent parameters respond to controlled perturbations of the formulation, reported in the same experimental studies [53, 54]. The first perturbation dilutes the liposome component by a factor of two prior to complexation (denoted '2x Dil'). The second varies the mRNA-to-liposome mass ratio from 1:6 (baseline) to 1:3 and 1:2, progressively increasing the relative mRNA content and thereby enhancing the driving force for mRNA-mediated aggregation.

Fixing the feature-dependent structure $Y(k) = \log(k)$ learned from the baseline condition, we re-estimate the dynamical parameters $r$ and $\tau$ independently for each experimental perturbation. The growth rate $r$ decreases under dilution and increases with higher mRNA loading (Fig. 11E), consistent with a reduced collision frequency at lower particle concentration and enhanced mRNA-mediated aggregation at higher mRNA loading, respectively. In contrast, $\tau$ remains consistent across perturbations (Fig. 11E), with values within the range observed for the baseline condition, suggesting that the steady-state cluster-size distribution is determined primarily by the geometry and affinity of the RNA–liposome interaction rather than by absolute concentration or stoichiometric ratio.

This example demonstrates several capabilities beyond the simulated systems: robustness to experimental noise and limited sampling, adaptive identification of a one-dimensional dynamical manifold from a 12-dimensional observable space, extraction of an interpretable power-law model without structural assumptions, and sensitivity of inferred parameters to physically meaningful perturbations. In contrast, a direct Smoluchowski fit for $k_{\max} = 12$ would require $\frac{1}{2}k_{\max}(k_{\max} + 1) = 78$ free parameters for only $T = 28$ time points, whereas our approach recovers a two-parameter dynamical model without specifying an evolution equation. These results show that the framework can extract mechanistically meaningful dynamics from realistic experimental data.

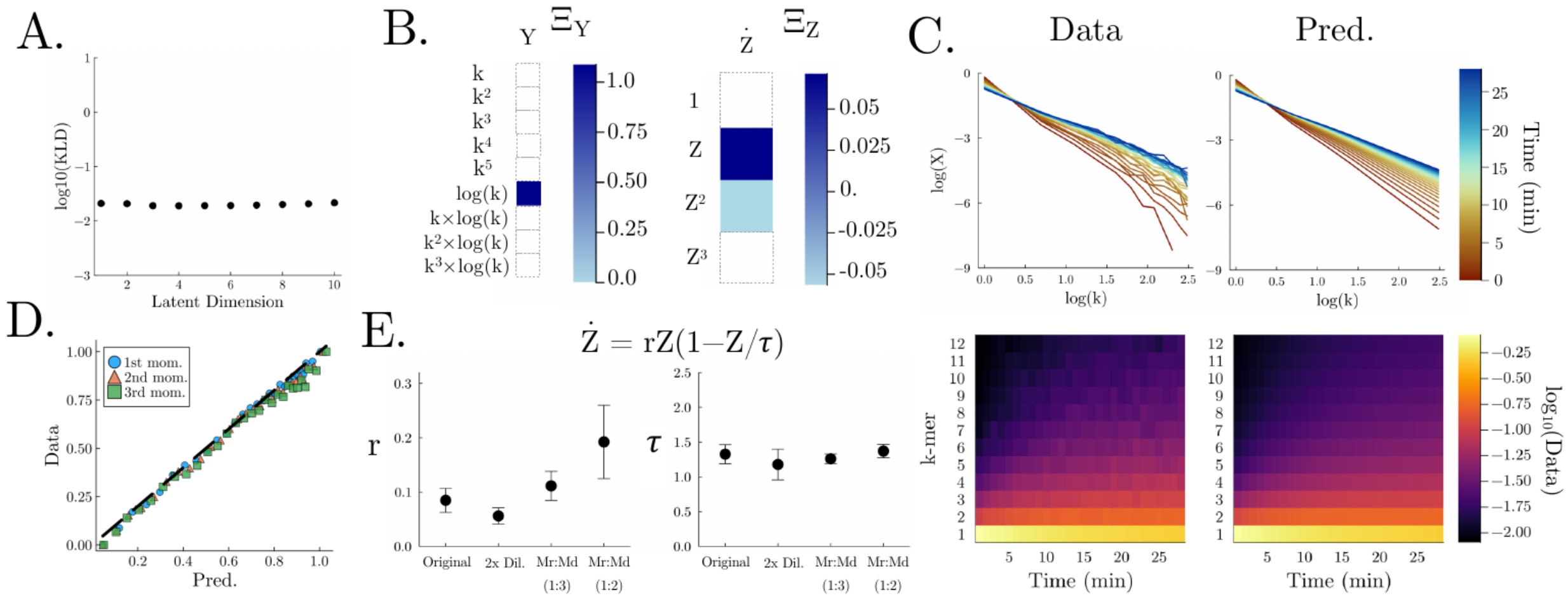

**Figure 11: Learned dynamics of experimental RNA-nanoparticle system. (A) Sweep of KLDs. (B) Learned models for time- and feature-dependent latents. (C) Data versus reconstruction. Top row is a log-log plot of the data and k-mer index. Bottom row is a heatmap for distribution over time. (D) Inferred power-law model accurately predicts several moments of the distribution, which were not constrained by the DME procedure. Data are min-max scaled according to the experimental data. (E) Coefficients of time-dependent model (shown above the plots) as a function of experimental perturbations. '2x Dil.' denotes a factor of two dilution of the liposome component prior to complexation with RNA and 'Mr:Md' denotes the mass ratio of RNA to liposomes. Data are presented as mean ± s.t.e. ($n = 7$ for the original formulation; $n = 3$ for all perturbation conditions).**

## Limitations

Here, we briefly identify some limitations of our work.

A restriction on data in *Edwin* is that the inputs, $p_{it}$, are required to be positively-valued and normalized. However, this restriction can be eased. For positively valued unnormalized data, we can adopt a modified KLD loss: $\mathcal{L}_{KLD}^{nn} = \sum_{it} p_{it} \log(p_{it}/q_{it}) - p_{it} + q_{it}$, as shown in [28]. As for the constraint of positively-valued data, though this covers a wide range of phenomena, inputs can be considered more generally as samples from an unknown distribution. In this case, a similar tack may be taken as in this work, but one can no longer use the same reconstruction loss, $\mathcal{L}_{KLD}$ in Eq. (11). Instead, this loss must be replaced with other strategies from generative modeling [79]. Given that our DME approximation is effectively an energy-based model, techniques such as contrastive divergence or score-matching may be suitable.

There are also several important considerations related to the model discovery component of this work. First, we only consider polynomial-based dynamical systems models. Though this comes with approximation power via the Weierstrass Theorem [80], it may not succinctly capture latent models composed of other analytical functions, such as sin,log, etc. This can be addressed via adding more complex libraries to our gradient calculations or via auto-differentiation by casting our Sparse Regression framework as a neural network. However, this may come with increased computational costs.

Second, recent work has shown that learning closed-form models from finite, noisy data exhibits a noise-driven learnability transition: below a critical noise level the true generating model is recoverable via probabilistic model selection, whereas above it simpler ("trivial") models become more parsimonious and the true model is unlearnable by any method [81]. Because SLIC performs information-theoretic model selection within a finite candidate library, *Edwin* is subject to the same fundamental tradeoff between noise level and model complexity. In high-noise regimes, simpler latent dynamics will necessarily be selected. The dynamic maximum entropy compression step may reduce effective noise by projecting data onto a structured low-dimensional manifold, but it does not eliminate the existence of a noise-dependent recoverability threshold.

Finally, other than the libraries themselves, we place no constraints on the models for the latents. Though the intention was to assume maximal ignorance of the underlying phenomena, one may possess information such as symmetries or invariants that may improve the quality of the underlying model. Enforcing such information may be performed as demonstrated in Ref. [18], which manifest themselves as affine constraints in the models, $\Xi_{Z,Y}$. Specifically, this is imposed via Lagrange multipliers in Eqs. (12) and/or (13): $\mathcal{L}_{Z,Y} \to \mathcal{L}_{Z,Y} + \lambda^{\top}(A\xi_{Z,Y} - b)$, where $\lambda$ is a vector of Lagrange multipliers, $\xi_{Z,Y}$ is a vectorized version of the models, $\Xi_{Z,Y}$, and $A, b$ encode the constraints. For example, in the 2D Brownian system, if we demand continuous rotational symmetry, which manifests itself as $Y(R\mathbf{x}) = Y(\mathbf{x})$, where $\mathbf{x} = (x, y)$ and $R$ is a rotational operator, then this condition is equivalent to $y\partial_x Y(x, y) = x\partial_y Y(x, y)$. If $Y(x, y) = \xi_1 x + \xi_2 y + \xi_3 x^2 + \xi_4 xy + \xi_5 y^2$, then this constraint is equivalent to $\xi_1 = \xi_2 = \xi_4 = 0$ and $\xi_3 = \xi_5$. In this case, the constraints may appear in matrix form as:

$$\begin{pmatrix} 1 & 0 & 0 & 0 & 0 \\ 0 & 1 & 0 & 0 & 0 \\ 0 & 0 & 1 & 0 & -1 \\ 0 & 0 & 0 & 1 & 0 \\ 0 & 0 & 0 & 0 & 0 \end{pmatrix} \begin{pmatrix} \xi_1 \\ \xi_2 \\ \xi_3 \\ \xi_4 \\ \xi_5 \end{pmatrix} = \mathbf{0}$$

However, for this simple case, it would be easier to simply exclude library terms that violate rotational symmetry.

Further, though our approach does not generally implicitly conserve quantities, we may enforce the dynamics of $Z$ to take a symplectic form: $\dot{Z} = J\nabla\mathcal{H}(Z)$, where $\mathcal{H}$ is the Hamiltonian of interest and $J$ is a skew-symmetric matrix ($J^\top = -J$). It is easily shown that $\dot{\mathcal{H}}(Z) = 0$. In this case, we may parameterize the Hamiltonian as before, $\mathcal{H}(Z) = \Theta_Z(Z)\Xi_Z$. More generally, it may be more satisfactory to adopt the ansatz espoused in [23]. One benefit of enforcing Hamiltonian structure is that it guarantees the stability of the latent dynamics, which has been the focus of recent work [82, 83]. As in these works, we may enforce stability by demanding $\dot{Z} = -M\nabla V(Z)$, where $M$ is positive semidefinite matrix; this results in $V(Z)$ being a Lyapunov function: $\dot{V}(Z) = -\nabla V^\top M \nabla V \leq 0$.

## Discussion

In this work, we present *Edwin*, a unified framework that simultaneously (i) compresses high-dimensional, nonnegative time-dependent data into low-dimensional latent coordinates $(\mathbf{Z}, \mathbf{Y})$ via the DME principle and (ii) learns interpretable sparse models $(\Xi_Z, \Xi_Y)$ governing the dynamics and feature dependence of those coordinates via our new SLIC algorithm. We validated this framework across a range of simulated and experimental systems, including stochastic diffusion, the Ornstein–Uhlenbeck process, self-assembling particles, spiking neural populations, low-rank recurrent neural networks, and a noisy experimental time series of aggregating RNA–liposome complexes. Across all systems, the method accurately reconstructs high-dimensional dynamics, recovers symbolic governing equations, and generalizes to unseen conditions with minimal additional data.

A central and distinguishing feature of our approach is the joint learning of time-dependent latents $\mathbf{Z}(t)$ and feature-dependent latents $\mathbf{Y}$, coupled via the exponential-family representation $p_{it} \propto \exp(-\sum_k Z_{tk} Y_{ik})$. Existing methods for joint compression and model discovery, such as SINDy-autoencoders, can recover symbolic equations for latent dynamics but leave the latent coordinates physically uninterpreted: their connection to observable quantities must be inferred post-hoc, if at all. In contrast, in *Edwin* the jointly learned model $\mathbf{Y} = \Theta_Y(\mathbf{x})\Xi_Y$ anchors each latent coordinate to external metadata $\mathbf{x}$, providing immediate physical meaning without additional analysis. In the Brownian diffusion example, this yields not only the dynamics $\dot{Z} \propto -Z^2$ but also the identification of $Z(t)$ as the inverse variance of the distribution; in the RNA–liposome system, it reveals a power-law scaling exponent whose value and perturbation response are interpretable in terms of known aggregation physics. This deeper interpretability is an architectural property of *Edwin* that cannot be recovered from reconstruction-only or dynamics-only compression schemes.

By utilizing a Maximum Entropy framework for compression, we not only gain the guarantees that our inferences are maximally unbiased [25, 26, 27], but by virtue of the connection with statistical mechanics and thermodynamics [25, 26, 27, 84], we gain access to the vast stochastic thermodynamics and thermodynamic optimal control literature [49, 85]. This permits us to then integrate our framework with control theory in a seamless manner. Moreover, via the principle of Maximum Entropy, there is a deep connection of statistical physics with networks [86], which may possess both complex dynamics and unknown low-dimensional structure [6], providing further motivation for the use of *Edwin*.

P.D.D. acknowledges R35GM142547 and J.G. acknowledges NIH grant R35GM146877 and Texas Global Faculty Research Seed Grant in support of this work.

## Description of systems

Here we give a detailed description of the simulated systems studied in this work. All simulations were performed in the Julia language using the DifferentialEquations.jl package [87].

*2D Brownian particle*. We simulate an isotropic Brownian process according to the stochastic ODE:

$$\begin{aligned} dx &= \sigma dW \\ dy &= \sigma dW \end{aligned}$$

Where $x, y$ denote the $x$ and $y$ spatial components, respectively, and $W$ is a Wiener process. We simulate 3000 trajectories from $t \in [0,10]$ with a timestep of 0.01. We then estimate the pdf of these trajectories at each timepoint using kernel density estimation with a Gaussian kernel with grids $x \in [-1,1]$ and $y \in [-1,1]$, with a grid spacing of 0.05. Thus, at each timepoint, we have a $41 \times 41$ matrix that we reshape into a column vector.

*OU process*. The OU process is a stochastic ODE of the form:

$$dx = \beta(\mu - x) + \sigma dW$$

Here, $W$ is a Wiener process. With $\beta = 0.4$ and $\mu = 0.7$, we simulate 3000 trajectories for $t \in [0,8]$ with a timestep of 0.01. As with the Brownian system, we then estimate the pdf at each timepoint using kernel density estimation along a grid $x \in [-1.5,1.5]$ with grid spacing 0.01.

*Self-assembling particles*. The equations for the self-assembling particles are the Smoluchowski equations [55]:

$$\frac{dn_k}{dt} = \frac{1}{2}\sum_{i+j=k} K_{ij}\, n_i n_j - n_k \sum_i K_{ik}\, n_i \quad (25)$$

Here $n_k$ denotes "k-mers", or clusters containing "$k$" particles and $K_{ij}$ denotes the kernel function, which is symmetric and dictates the rate at which different clusters assemble. The first term on the RHS of the above expression determines the ways in which one can create k-mers via the assembly of smaller clusters; the second term determines how k-mers are destroyed via merging with other clusters.

We consider the case in which the kernel $K_{ij}$ is a constant, $K$; this approximation is commonly used as an approximation for self-assembly dominated by Brownian motion [55, 77]. We simulate this system with a max cluster size of 100, for $t \in [0,100]$ with a timestep of 0.05.

*Neural population*. In this final system, we simulate two, two-dimensional linear differential equations, $\dot{X} = AX$, each possessing matrices $A$ with different spectra. This leads to different dynamical behaviors for each system simulated. Each nonzero element was chosen randomly, with the constraint that the dynamics must be stable (i.e., the real part of the eigenvalues are negative). The first operator, $A_1$ has only complex eigenvalues, and thus results in oscillatory behavior.

For each dynamical system, we generate 10 randomly-chosen initial conditions and simulate the trajectories for each for $t \in [0,8]$ with a timestep of 0.01. Each of these initial conditions is then embedded into a firing rate $\lambda_{tn} = \exp(b_n + \sum_k C_{nk}\, X_{tk})$, where $t$ denotes the time index, $k$ denotes the latent index, and $n$ denotes the neuron; $C_{nk}$ and $b_n$ are sampled randomly from a Gaussian normal distribution. The number of neuron spikes within a time $dt$ is then sampled from a Poisson distribution, $Pois(\lambda_{tn} dt)$. For each initial condition of the latent dynamics, we simulate 100 neuron firing trials, which is meant to mimic experimental repeats; these trials are then averaged. Our final data matrix is $100 \times 801 \times 10$.

## The SLIC Model Discovery Algorithm

### *Algorithm Overview*

In this section, we give a brief overview of the most essential component of our SLIC algorithm solving the regression problem outlined in Eq. (5), reproduced here for convenience:

$$||\mathbf{Y} - \Theta(\mathbf{X})\Xi||_2^2 + \nu^2 ||\Xi||_0 \quad (26)$$

Inputs: $\mathbf{Y}$, $\Theta$ Outputs: $\Xi$ $\Xi \leftarrow \Theta^\dagger \mathbf{Y}$ $score \leftarrow SLIC(\Xi)$ $iter \leftarrow 1$ $vs \leftarrow nonzero\ elements\ of\ |\Xi|$ $\Xi_{new} \leftarrow copy(\Xi)$ $inds \leftarrow |\Xi| < \nu$ $0 \leftarrow \Xi_{new}(inds)$ $\Xi_{new} \leftarrow \Theta^\dagger \mathbf{Y}$ $new\ score \leftarrow SLIC(\Xi_{new})$ $score \leftarrow new\ score$ $\Xi \leftarrow \Xi_{new}$ $iter \leftarrow iter + 1$

In Algorithm 1, $SLIC(\Xi) = n\log(\hat{\epsilon}k)$, where $\hat{\epsilon} = \frac{1}{n}||Y - \Theta\Xi||_2^2$ and $k$ is the number of free parameters in the model, $||\Xi||_0 + 1$, which includes the unknown variance of the noise distribution. The † denotes a matrix pseudoinverse.

To quantify the computational advantage of SLIC over evolutionary symbolic regression, we compare mean runtimes against PySR [8] on the four nonlinear ODE systems introduced in Appendix 13, across dataset sizes of 1,000, 5,000, and 10,000 data points. As shown in Fig. S2, SLIC is approximately three to four orders of magnitude faster than PySR across all systems and dataset sizes, while recovering equivalent governing equations (Fig. 2). This speedup reflects two complementary factors: SLIC operates within a fixed candidate library rather than performing combinatorial search over an open expression space, and model selection reduces to computing the pseudoinverse of the library matrices, which is computationally cheap once the active set is identified. Together, these properties make SLIC tractable to integrate into the joint DME optimization scheme.

### *A toy example for Algorithm [alg:slic]*

To illustrate selection behavior, consider the following toy example. All numerical coefficient values given here are fictitious, and are purely for illustrative purposes. Let $\mathbf{X}, \mathbf{Y} \in \mathbb{R}^{n\times 1}$, which we simply denote as $x$ and $y$, respectively. Assume they have the following mathematical relationship:

$$y = x^2 + \varepsilon$$

with $\varepsilon \sim \mathcal{N}(0, \sigma^2)$.

Let $\Theta(x) = [1\ x\ x^2\ x^3]$, meaning that we approximate $y$ as $\hat{y} = \Theta(x)\Xi$, or:

$$\hat{y} = \xi_1 + \xi_2 x + \xi_3 x^2 + \xi_4 x^3$$

Our goal is to estimate the $\xi_i$.

The first step of Algorithm [alg:slic] is to naively estimate the unknown coefficients by minimizing $||y - \Theta(x)\Xi||_2^2$. This gives an initial, non-sparse estimate:

$$\hat{y} = -0.001 + 0.03x + 0.90x^2 + 0.1x^3$$

This model is then scored with SLIC. The magnitude of the coefficients, $[0.001, 0.03, 0.1, 0.9]$, becomes the array of thresholds, $vs$; we may neglect the largest $\nu = 0.9$, as this results in a trivial model. Each threshold, $\nu \in vs$, results in a new model as a result of pruning all terms whose magnitude falls beneath it:

$$\begin{aligned} \hat{y}_1 &= -0.002x + 0.90x^2 + 0.01x^3 \\ \hat{y}_2 &= 0.001x + 0.95x^2 \\ \hat{y}_3 &= 1.01x^2 \end{aligned}$$

Note that after pruning, all nonzero entries are updated. These models are then all scored by SLIC:

$$\hat{y}_1 \rightarrow \text{SLIC}_1$$
$$\hat{y}_2 \rightarrow \text{SLIC}_2$$
$$\hat{y}_3 \rightarrow \text{SLIC}_3$$

The model among these with the smallest negative $\Delta\text{SLIC}_i = \text{SLIC}_i - \text{SLIC}_{best}$, where $\text{SLIC}_{best}$ is the current best model, is taken as the new best model; this new best model is then used to generate a new array of $\nu s$ for the next iteration.

### *The Rationale for SLIC*

Every algorithm possesses hyperparameters, parameters that are not learned by the algorithm during the training process. Therefore, the process of model selection, in which one must somehow choose the "best" set of these hyperparameters, is of the utmost importance.

Of the many model selection criteria [88], those that use information-based scoring criteria [45] are particularly attractive in model discovery applications, as they explicitly contain penalties for complex models. The majority of these information criteria (denoted as $IC$), under the assumption of Gaussian-distributed errors, take the following form [45]:

$$IC(n,k) = n\log(\hat{\epsilon}) + s(n,k) \tag{27}$$

Where $n$ is the number of data points, $k$ is the model complexity (the number of free parameters), and $\hat{\epsilon}$ is the mean-squared error between the data and the model; $s(n,k)$ is a function that penalizes the complexity of the model. The first term in Eq. (27) penalizes models that fit poorly. The "best" model is the one that has the smallest overall $IC(n,k)$.

Interestingly, essentially all information-based model selection criteria possess $s(n,k)$ that scale sublinearly with $n$. Therefore, as $n \rightarrow \infty$, the first term in Eq. (27), which scales linearly in $n$, is going to dominate the sum. Though this may be sensible for some applications, it is clearly not desirable for our purposes, where we assume the model we would like to select is invariant to the amount of data *Edwin* processes.

Therefore, we argue that $s(n,k)$ should scale linearly with $n$: $s(n,k) \sim n\log(f(k))$. As a starting point, we take this to be an equality. More generally, we could have the function $f$ also be a function of $n$ that asymptotically approaches $f(k)$, but we neglect such cases.

To obtain a functional form for $f(k)$, we let $IC(n,k) = F(\hat{\epsilon},k) = n\log(\hat{\epsilon}) + n\log(f(k)) = n\log(\hat{\epsilon}f(k))$. Now, we assume equitable penalization for fit and complexity, or $F(\hat{\epsilon},k) = F(k,\hat{\epsilon})$. This then implies:

$$\begin{aligned} \hat{\epsilon}f(k) &= kf(\hat{\epsilon})) \\ \frac{f(k)}{k} &= \frac{f(\hat{\epsilon})}{\hat{\epsilon}} \end{aligned} \tag{28,29}$$

Assuming the approximate independence of these variables, the only way for Eq. (29) to hold is for the ratio to be constant, or $f(k) \propto k$. Given that we only care about differences in $IC$ for the sake of model selection, we can take this to be an equality. Thus, we arrive at our functional form for SLIC, $n\log(\hat{\epsilon}k)$.

### *Galerkin Projection Scheme*

When we are concerned with discovering a sparse model of a dynamical system, which is the case for our latents $Z$, the $Y$ in Eq. (26) is a time derivative of state variables, $\dot{X}$. Unfortunately, we often do not have access to this quantity and therefore must estimate it from data. Although there exist many such procedures for doing so [89], these procedures often introduce additional noise into our optimization problem, which complicates model discovery.

However, by transforming our problem into the so-called "weak form", which amounts to integrating by parts, we can avoid numerically differentiating altogether, thus dramatically improving noise robustness [15]. For illustrative purposes, consider our one-dimensional problem:

$$\dot{x} = \Theta(x)\Xi \tag{30}$$

Where, as previously stated, $\Theta(x)$ is a library of nonlinearities and $\Xi$ is our matrix of coefficients to be determined. Let $\langle f \rangle_w = \int_{t_1}^{t_2} d\,t f(t)w(t)$. Now, if we assume Eq. (30) holds, then:

$$\langle \dot{x} \rangle_w = \langle \Theta(x)\Xi \rangle_w \tag{31}$$

Now, assuming $w(t_1) = w(t_2) = 0$ and integrating by parts, we have:

$$\langle x \rangle_{-\dot{w}} = \langle \Theta(x)\Xi \rangle_w \tag{32}$$

Thus, we have transferred numerical differentiation of our data $x$ to the derivative of a known function, $w(t)$. Given the linearity of integration, we may express both sides, in matrix form as:

$$DX = W\Theta(X)\Xi \tag{33}$$

In practice, we choose $w(t) = C[(t_2 - t)(t - t_1)]^p$, where $C$ is a constant chosen such that $\langle 1 \rangle_w = 1$, and build the matrices $D, W$ via trapezoidal integration.

We note that the use of the weak form to avoid numerical differentiation is shared with the Weak SINDy method of [15]. The key algorithmic distinction lies in how sparsity is enforced and model selection is performed. In Weak SINDy, the practitioner must specify a grid of pruning thresholds $\nu$ and evaluate candidate models across this grid. In *Edwin*, the candidate thresholds are generated automatically from the magnitude of the current model coefficients, each candidate is scored using SLIC, and the optimal model is selected without user intervention. This automation is essential for our framework, in which model selection must occur repeatedly and without manual input throughout the joint DME optimization.

## The Primary Algorithm in this Work

### *Algorithm Overview*

Here we provide pseudocode for *Edwin*; a schematic is given in Fig. 12.

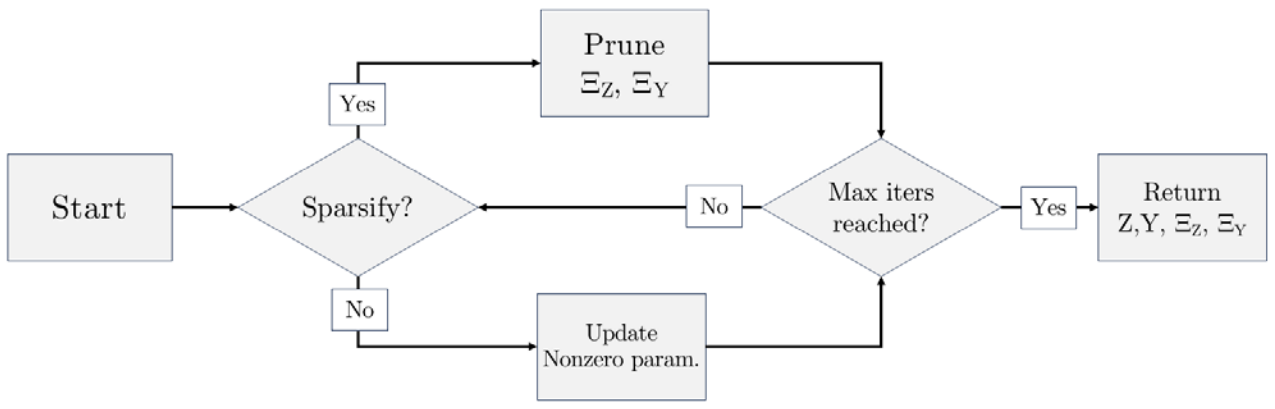

**Figure 12: Flow schematic for Algorithm [alg:main].**

Inputs: $\mathbf{P}$, $\Theta_Z$, $\Theta_Y$, $K$, $\eta$, $maxiters$, $M$ Outputs: $\mathbf{Z}$, $\mathbf{Y}$, $\Xi_Z$, $\Xi_Y$ $iter \leftarrow 1$ $\mathbf{Z} \leftarrow \mathbf{Z} - \eta \frac{\partial \mathcal{L}}{\partial \mathbf{Z}}$ $\mathbf{Y} \leftarrow \mathbf{Y} - \eta \frac{\partial \mathcal{L}}{\partial \mathbf{Y}}$ $\Xi_Z \leftarrow SLIC(Z, \Theta_Z)$ $\Xi_Y \leftarrow SLIC(Y, \Theta_Y)$ $\Xi_Z \leftarrow \Theta_Z^\dagger(\mathbf{Z})\dot{\mathbf{Z}}$ $\Xi_Y \leftarrow \Theta_Y^\dagger(\mathbf{x})\mathbf{Y}$ $iter \leftarrow iter + 1$

In Algorithm 2, $\partial \mathcal{L}/\partial \mathbf{Z}$ and $\partial \mathcal{L}/\partial \mathbf{Y}$ denote partial derivatives of the loss function with respect to $\mathbf{Z}$ and $\mathbf{Y}$, respectively. We compute them directly and update the unknown latents using gradient descent, with $\eta$ as the learning rate, set by default to 0.001. $M$ is set to 50,000 iterations and $maxiters$ to 150,000. We discuss how these gradients are computed in the following section.

The $\mathbf{Z}, \mathbf{Y}$ for each system is initialized to the first $K$ columns of the singular value decomposition. Specifically, if $\mathbf{P} = U\Sigma V^\dagger$, then we initialize the values of $\mathbf{Z}$ and $\mathbf{Y}$ with the first $K$ columns of $U$ and $V$, respectively.

### *Hyperparameter Selection*

The loss function in Eq. (9) contains two regularization hyperparameters, $\lambda_{\dot{Z}}$ and $\lambda_Y$, which control the relative strength of the dynamical and feature model terms with respect to the reconstruction loss. To select these automatically, we perform a grid search over $\lambda_{\dot{Z}}, \lambda_Y \in \{10^{-1}, 10^0, 10^1\}$ independently, yielding nine candidate combinations. For each combination, we run the full *Edwin* optimization and compute the combined SLIC score of the resulting models. The final hyperparameters are chosen as the combination that minimizes the sum of the SLIC scores of $\Xi_Z$ and $\Xi_Y$, reflecting the joint parsimony of both discovered models.

### *Gradients of the Loss Function*

When our library consists of polynomials, we can compute the gradients of the loss function directly without resorting to auto-differentiation software. Recall that our loss function in Eq. (9) is:

$$\mathcal{L} = \mathcal{L}_{KLD}(\mathbf{Z}, \mathbf{Y}) + \lambda_{\dot{Z}}\mathcal{L}_{\dot{Z}}(\mathbf{Z}, \Xi_Z) + \lambda_Y\mathcal{L}_Y(\mathbf{Y}, \Xi_Y)$$

With each of the loss terms:

$$\begin{aligned} \mathcal{L}_{KLD}(\mathbf{Z}, \mathbf{Y}) &= \sum_{i,t} p_{it} \log\left(\frac{p_{it}}{q_{it}}\right) \\ &= \sum_{i,t,k} p_{it}\, Z_{tk} Y_{ik} + \sum_t \log \Omega_t \end{aligned} \quad (34,35)$$

$$\begin{aligned} \mathcal{L}_{\dot{Z}}(\mathbf{Z}, \Xi_Z) &= \frac{1}{2}||\dot{\mathbf{Z}} - \Theta_Z(\mathbf{Z})\Xi_Z||_2^2 + \nu_Z^2||\Xi_Z||_0 \\ &= \frac{1}{2}||D\mathbf{Z} - W\Theta_Z(\mathbf{Z})\Xi_Z||_2^2 + \nu_Z^2||\Xi_Z||_0 \end{aligned} \quad (36,37)$$

$$\mathcal{L}_Y(\mathbf{Y}, \Xi_Y) = \frac{1}{2}||\mathbf{Y} - \Theta_Y(\mathbf{x})\Xi_Y||_2^2 + \nu_Y^2||\Xi_Y||_0 \quad (38)$$

Where $D$ and $W$ in Eq. (37) are matrices from our Galerkin projection scheme Eq. [D&W].

Given that we solve for $\Xi_Y$ and $\Xi_Z$ via regression (no gradient descent), we need only determine $\partial \mathcal{L}/\partial Y_{ik}$ and $\partial \mathcal{L}/\partial Z_{tk}$. The derivative of Eq. (35) with respect to these quantities is simply:

$$\frac{\partial \mathcal{L}_{KLD}}{\partial Z_{tk}} = \sum_i (p_{it} - q_{it})Y_{ik} \quad (39)$$

$$\frac{\partial \mathcal{L}_{KLD}}{\partial Y_{ik}} = \sum_t (p_{it} - q_{it})Z_{tk} \quad (40)$$

Further, $\partial \mathcal{L}_Y/\partial Y_{ik}$ is straightforward, yielding, for the total gradient:

$$\frac{\partial \mathcal{L}}{\partial Y_{ik}} = \sum_t (p_{it} - q_{it})Z_{tk} + \lambda_Y(Y_{ik} - [\Theta_Y\Xi_Y]_{ik}) \quad (41)$$

In matrix form, this reads:

$$\frac{\partial \mathcal{L}}{\partial \mathbf{Y}} = (\mathbf{P} - \mathbf{Q})\mathbf{Z} + \lambda_Y(\mathbf{Y} - \Theta_Y\Xi_Y) \quad (42)$$

Where $\mathbf{Q}_{it} := q_{it}$. Thus, the first term on the RHS above penalizes poor data reconstruction and the second term penalizes $\mathbf{Y}$'s that deviate from the inductive bias we place on these feature-dependent latents.

The more challenging computation is that of $\partial \mathcal{L}_Z/\partial Z_{tk}$, as a result of the nonlinear dependence of $\Theta_Z(\mathbf{Z})$ on $\mathbf{Z}$. To analytically determine the gradient, we first decompose the first term on the RHS of Eq. (37) into three terms using $||A||_2^2 = tr(A^\top A)$:

$$\begin{aligned} &tr(\mathbf{Z}^\top D^\top D\mathbf{Z}) - 2tr(\mathbf{Z}^\top D^\top W\Theta_Z(\mathbf{Z})\Xi_Z) \\ &+tr(\Xi_Z^\top \Theta_Z(\mathbf{Z})^\top W^\top W\Theta_Z(\mathbf{Z})\Xi_Z^\top) \end{aligned} \quad (43,44)$$

For notational simplicity, we let $\Theta = \Theta_Z(\mathbf{Z})$ and $\Xi = \Xi_Z$. The derivative of the first term on the RHS is straightforward, and is $2D^\top D\mathbf{Z}$. Thus, we tackle differentiating the second and third terms on the RHS. We can write the second trace term on the RHS in matrix form as (excluding the factor of $-2$):

$$tr(\mathbf{Z}^\top D^\top W\Theta_Z(\mathbf{Z})\Xi_Z) = \sum_{ijklm} Z_{ij}^\top D_{jk}^\top W_{kl}\Theta_{lm}\Xi_{mi} \quad (45)$$

The derivative of this term w.r.t $Z_{sn}$ is:

$$\sum_{ijklm} \delta_{js}\, \delta_{in} D_{jk}^\top W_{kl}\Theta_{lm}\Xi_{mi} + \sum_{ijklm} Z_{ij}^\top D_{jk}^\top W_{kl} \frac{\partial \Theta_{lm}}{\partial Z_{sn}} \Xi_{mi} \quad (46)$$

The first term above simplifies to $(D^{\top} W \Theta \Xi)_{sn}$. To achieve a closed form for the second term, we assume that our library is polynomial: $\Theta_{lm} = \prod_k Z_{lk}^{N_{mk}}$, where $N_{mk}$ are the polynomial degrees. Thus, for each timepoint (tracked by the index $l$), the $m$th library term is simply a product of the latents $Z_{lk}$ ($k \in [1, K]$) at that timepoint. With this assumption,

$$\frac{\partial \Theta_{lm}}{\partial Z_{sn}} = \Theta_{lm} N_{mn} \delta_{sl} \frac{1}{Z_{ln}} \tag{47}$$

Substituting this into the second term of Eq. (46) and simplifying:

$$\sum_{ijkm} Z_{ij}^{\top} D_{jk}^{\top} W_{ks} \left(\Theta_{sm} N_{mn} \frac{1}{Z_{sn}}\right) \Xi_{mi} \tag{48}$$

Now, the indices of the terms being summed over must form a chain going from $s \to n$ to find a closed-form matrix expression. This can be achieved via:

$$\sum_{ijkm} W_{sk}^{\top} D_{kj} Z_{ji} \Xi_{im}^{\top} \Theta_{sm} N_{mn} \frac{1}{Z_{sn}} \tag{49}$$

Summing over indices $i, j, k$, we get:

$$\sum_m (W^{\top} D Z \Xi^{\top})_{sm} \Theta_{sm} N_{mn} \frac{1}{Z_{sn}} \tag{50}$$

Now, the first two terms can be expressed using the Hadamard Product [90], defined as $(A \odot B)_{ij} = A_{ij} B_{ij}$. This is simply element-wise multiplication, and is commonly used in most programming languages. Doing this:

$$\sum_m (W^{\top} D Z \Xi^{\top} \odot \Theta)_{sm} N_{mn} \frac{1}{Z_{sn}} \tag{51}$$

Finally, summing over $m$ and using the Hadamard Product again, we get:

$$\left(((W^{\top} D Z \Xi^{\top} \odot \Theta) N) \odot \frac{1}{Z}\right)_{sm} \tag{52}$$

Where $N$ is the matrix whose elements are the polynomial orders, $N_{mn}$, and $1/Z$ denotes the matrix whose elements are $1/Z_{sn}$, which is just element-wise inversion of $Z$.

Similar algebra can be performed for the final term in Eq. (44), which yields:

$$\left(((W^{\top} W \Theta \Xi \Xi^{\top} \odot \Theta) N) \odot \frac{1}{Z}\right)_{sm} \tag{53}$$

Letting $\Delta_Z = DZ - W\Theta_Z \Xi_Z$, we have for the total gradient:

$$\frac{\partial \mathcal{L}}{\partial \mathbf{Z}} = (\mathbf{P} - \mathbf{Q})^{\top} \mathbf{Y} + \lambda_Z \left(D^{\top} \Delta_Z - ((W^{\top} \Delta_Z \Xi^{\top} \odot \Theta_Z) N) \odot \frac{1}{Z}\right)$$

*Computational Complexity Analysis*

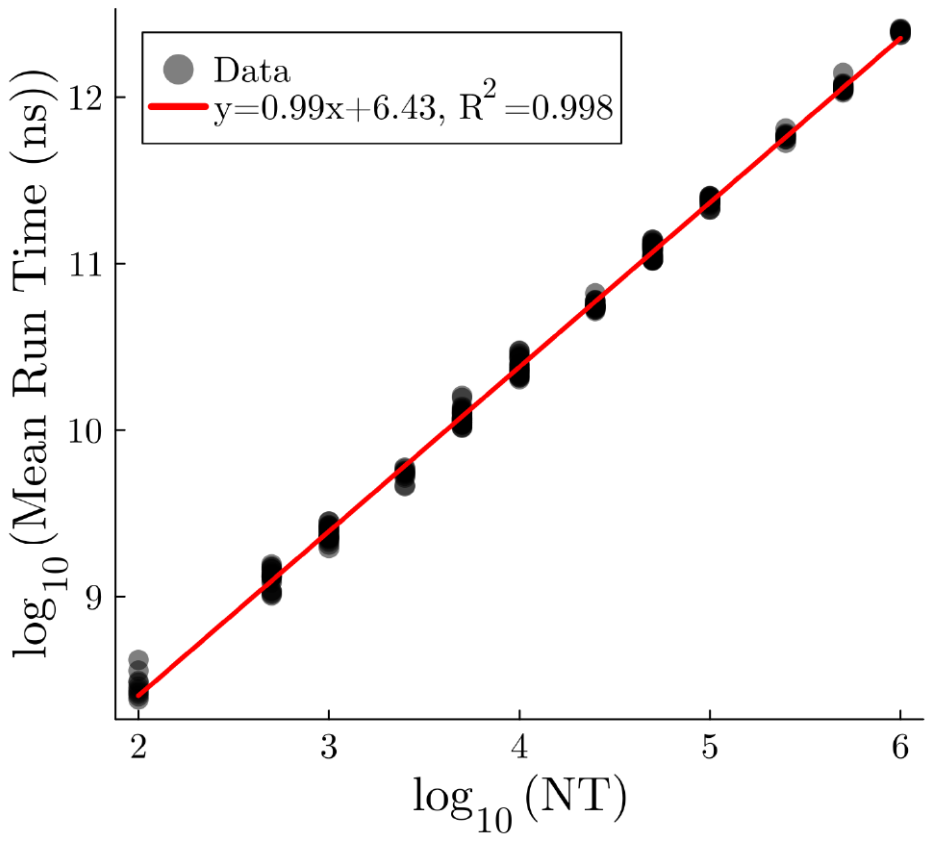

**Figure 13: Runtime scaling of the primary algorithm (Algorithm [alg:main]) as a function of the product $NT$. Each point is the mean over ten independent runs; $K$, swept from 1 to 10, contributes to the constant prefactor rather than the asymptotic scaling, consistent with the dominant $\mathcal{O}(INTK)$ complexity.**

Let $N$ denote the feature dimension (rows of $P$), $T$ the number of timepoints, $K$ the latent dimension, $l_Z$ and $l_Y$ the number of candidate library terms in $\Theta_Z$ and $\Theta_Y$, $s_Z$ and $s_Y$ the number of active (nonzero) terms after sparsification, $I$ the total number of gradient iterations, and $M$ the sparsification interval. The data and latent variables have dimensions

$$P \in \mathbb{R}^{N \times T}, \quad Z \in \mathbb{R}^{T \times K}, \quad Y \in \mathbb{R}^{N \times K}.$$

Throughout, we assume $K \ll \min(N, T)$ and $s_Z \ll l_Z$, $s_Y \ll l_Y$ after sparsification.

Each gradient iteration consists of three main computational components: DME reconstruction and gradient evaluation, regression updates of active coefficients, and occasional sparsification via SLIC. The reconstruction step requires computing the matrix product $ZY^{\top}$ in order to evaluate

$$q_{it} = \frac{1}{\Omega_t} \exp\left(-\sum_{k=1}^{K} Z_{tk} Y_{ik}\right),$$

which costs $\mathcal{O}(NTK)$. The elementwise exponential and normalization contribute an additional $\mathcal{O}(NT)$ operations, so the dominant reconstruction and gradient cost per iteration is $\mathcal{O}(NTK)$.

When sparsification is not performed, regression updates are restricted to the active terms. Evaluating the reduced libraries $\Theta_Z(Z) \in \mathbb{R}^{T \times s_Z}$ and $\Theta_Y(Y) \in \mathbb{R}^{N \times s_Y}$ costs $\mathcal{O}(Ts_Z)$ and $\mathcal{O}(Ns_Y)$, respectively. Solving the associated least squares problems via QR decomposition or normal equations scales as $\mathcal{O}(Ts_Z^2 + s_Z^3)$ for $Z$ and $\mathcal{O}(Ns_Y^2 + s_Y^3)$ for $Y$. Under typical sparsity assumptions, these terms remain lower order compared to $\mathcal{O}(NTK)$.

Every $M$ iterations, SLIC performs iterative thresholding over candidate coefficients. In the worst case, if $l$ candidate terms are present and $n$ samples are used, the sparsification step requires up to $\mathcal{O}(nl^2 + l^3)$ operations per regression solve. Once the algorithm has converged and the active set has been reduced to $s \ll l$ terms, the per-call cost decreases to $\mathcal{O}(ns^2 + s^3)$. Since sparsification is invoked every $M$ iterations, the steady-state sparsification cost over training scales as

$$\mathcal{O}\left(\frac{I}{M}(Ts_Z^3 + Ns_Y^3)\right).$$

Combining all contributions, the overall training complexity is

$$\mathcal{O}(INTK) + \mathcal{O}(I(Ts_Z^2 + Ns_Y^2)) + \mathcal{O}\left(\frac{I}{M}(Ts_Z^3 + Ns_Y^3)\right).$$

In typical scientific applications where $N, T \gg K, s_Z, s_Y$, the dominant term is $\mathcal{O}(INTK)$, implying linear scaling in the data size $NT$ and latent dimension $K$, with cubic dependence only on the small number of active symbolic terms. This is confirmed empirically by generating random, positively-valued matrices with $N, T \in [10,50,100,500,1000]$, $K \in [1,10]$ and running the algorithm ten times, taking the mean of the resulting runtime (Fig. 13). Memory usage is dominated by storing $P$, which requires $\mathcal{O}(NT)$ storage, with additional lower-order contributions from storing $Z$, $Y$, and the library matrices.

## On DME and Maximum Caliber (MaxCal)

The Principle of Maximum Caliber (MaxCal) is often considered the time-dependent generalization of The Principle of Maximum Entropy (MaxEnt) [26, 27]. If $\Gamma$ denotes a "trajectory" $i_1, i_2, \dots, i_T$, where $i_t$ is a feature $i$ at time $t$, then the objective of MaxCal is to minimize the following with respect to $p_\Gamma$:

$$-\sum_\Gamma p_\Gamma \log p_\Gamma + \sum_k Z_k \left(\langle Y_{\Gamma,k}\rangle - \sum_\Gamma p_\Gamma Y_{\Gamma,k}\right) \tag{54}$$

Where $Y_{\Gamma,k}$ denotes the $k$-th observable ($k \in [1, K]$) over $\Gamma$ and $Z_k$ is a Lagrange multiplier. The solution to minimizing the above with respect to $p_\Gamma$ is: $p_\Gamma \propto \exp\left(-\sum_k Z_k Y_{\Gamma,k}\right)$.

In this work, in the above notation, we used the DME principle:

$$p_{it} \propto \exp(-\sum_k Z_{tk} Y_{ik}) \tag{55}$$

The natural question is: if MaxCal is the general time-dependent extension of MaxEnt, how is DME derived from MaxCal?

The idea is straightforward. Our $K$ constraints are not taken over the entire path, but only at each timepoint where we collect data:

$$Z_{11}\left(\langle Y_{i_1,1}\rangle - \sum_\Gamma p_\Gamma Y_{i_1,1}\right) + \cdots$$
$$Z_{1K}\left(\langle Y_{i_1,K}\rangle - \sum_\Gamma p_\Gamma Y_{i_1,K}\right) + \cdots$$
$$Z_{TK}\left(\langle Y_{i_T,K}\rangle - \sum_\Gamma p_\Gamma Y_{i_T,K}\right)$$

These constraints can be written as:

$$\sum_{\tau,k} Z_{\tau k}\left(\langle Y_{i_\tau,k}\rangle - \sum_\Gamma p_\Gamma Y_{i_\tau,k}\right) \tag{56}$$

Then, the solution for the distribution is:

$$\begin{aligned} p_\Gamma &\propto \exp\left(-\sum_{\tau,k} Z_{\tau k} Y_{i_\tau,k}\right) \\ &= \prod_\tau \exp\left(-\sum_k Z_{\tau k} Y_{i_\tau,k}\right) \end{aligned}$$

Thus, marginalizing over all $\tau \neq t$, we get:

$$p_{i_t} = p_{it} \propto \exp\left(-\sum_k Z_{tk} Y_{i_t,k}\right) \tag{57}$$

Finally, if we assume $Y_{i_t,k} = Y_{ik}$ for all $t$, then we recover Eqn. (55). Thus, DME emerges as a minimal MaxCal principle in which one: (1) constrains some set of observables only over the available timepoints and (2) the observables are independent of time.

## Dynamic Nonnegative Matrix Factorization

Dynamic Nonnegative Matrix Factorization (Dyn-NMF) extends standard nonnegative matrix factorization by incorporating temporal dynamics into the latent activations through a state-space formulation [66]. Given a nonnegative data matrix $\mathbf{X} \in \mathbb{R}_+^{K\times T}$, Dyn-NMF seeks a factorization

$$\mathbf{X} \approx \mathbf{W}\mathbf{H} \tag{58}$$

where $\mathbf{W} \in \mathbb{R}_+^{K\times I}$ is a nonnegative basis matrix and $\mathbf{H} \in \mathbb{R}_+^{I\times T}$ contains time-dependent latent activations. Each observation $\mathbf{x}_t$ is modeled as a nonnegative linear combination of basis vectors,

$$\mathbf{x}_t \approx \mathbf{W}\mathbf{h}_t \tag{59}$$

with the latent state constrained to lie on the probability simplex,

$$h_{i,t} \geq 0, \qquad \sum_{i=1}^{I} h_{i,t} = 1 \tag{60}$$

This normalization corresponds to the probabilistic latent component analysis (PLCA) interpretation used in the implementation.

Temporal structure is introduced via an autoregressive model of order $J$ acting on the latent variables,

$$\mathbf{h}_t \approx \sum_{j=1}^{J} \mathbf{A}_j \, \mathbf{h}_{t-j} \tag{61}$$

where $\mathbf{A}_j \in \mathbb{R}_+^{I\times I}$ are nonnegative transition matrices. The predicted latent state

$$\boldsymbol{\eta}_t = \sum_{j=1}^{J} \mathbf{A}_j \, \mathbf{h}_{t-j} \tag{62}$$

serves as a dynamical prior for $\mathbf{h}_t$ during inference.

Model parameters are estimated by alternating updates of $\mathbf{W}$, $\mathbf{H}$, and $\{\mathbf{A}_j\}_{j=1}^{J}$ using multiplicative and expectation–maximization–type rules that preserve nonnegativity. Responsibilities describing the contribution of component $i$ to feature $k$ at time $t$ are given by

$$\Gamma_{kit} = \frac{W_{ki} h_{it}}{\sum_{i'} W_{ki'} h_{i't}} \tag{63}$$

The basis matrix is updated by aggregating contributions across time,

$$W_{ki} \propto \sum_{t=1}^{T} x_{kt} \, \Gamma_{kit} \tag{64}$$

followed by column normalization. Latent activations $\mathbf{h}_t$ are updated using both the observation $\mathbf{x}_t$ and the dynamical prior $\boldsymbol{\eta}_t$, subject to the simplex constraint. The transition matrices are estimated by fitting the autoregressive relation

$$\mathbf{H} \approx \sum_{j=1}^{J} \mathbf{A}_j \, \mathbf{H}^{(j)} \tag{65}$$

where $\mathbf{H}^{(j)}$ contains time-shifted latent vectors.

After training, future latent states are predicted recursively using the learned dynamics,

$$\hat{\mathbf{h}}_t = \sum_{j=1}^{J} \mathbf{A}_j \, \hat{\mathbf{h}}_{t-j} \tag{66}$$

and predicted observations are obtained via

$$\hat{\mathbf{x}}_t = \mathbf{W}\hat{\mathbf{h}}_t \tag{67}$$

Nonnegativity and normalization are enforced at each step to maintain consistency with the PLCA formulation.

## Description of encoder-based architectures

*Standard autoencoder (AE)*. Let $\varphi$ be a neural-network encoder that maps an input $\mathbf{x} \in \mathbb{R}^N$ to a latent space $\mathbf{z} \in \mathbb{R}^K$ (with $K < N$) and $\psi$ be a neural-network decoder that maps $\mathbf{z}$ to a reconstructed input $\hat{\mathbf{x}} \in \mathbb{R}^N$, i.e., $\hat{\mathbf{x}} = \psi(\varphi(\mathbf{x}))$. The loss function is the squared error between the data and the reconstruction: $\mathcal{L} = ||\mathbf{x} - \hat{\mathbf{x}}||_2^2$.

*SINDy-autoencoders (SINDy-AE)*. As in the standard AE, SINDy-AEs [22] possess an encoder, $\varphi$, and a decoder, $\psi$. The goal here, however, is not to merely learn a compression scheme, but also a model for the latent dynamics: $\dot{\mathbf{z}} = \Theta(\mathbf{z})\Xi$; here a sparse $\Xi$ is learned from the data. Thus, the loss generally consists of several terms. First, there is a reconstruction loss: $\mathcal{L}_{rec} = ||\mathbf{x} - \hat{\mathbf{x}}||_2^2$. Then, there is loss for the latent space model: $\mathcal{L}_{mod} = ||\dot{\mathbf{z}} - \Theta(\mathbf{z})\Xi||_2^2$; an additional $L_1$ regularization loss for the dynamics model is used: $\mathcal{L}_{reg} = ||\Xi||_1$. Finally, there is a loss that enforces reconstruction of the rate of change of the input from the dynamics of the latent: $\mathcal{L}_{rate} = ||\dot{\mathbf{x}} - \dot{\mathbf{z}} \cdot \nabla\psi(\mathbf{z})||_2^2$. Thus, together:

$$\mathcal{L} = \mathcal{L}_{rec} + \lambda_1\mathcal{L}_{mod} + \lambda_2\mathcal{L}_{rate} + \lambda_3\mathcal{L}_{reg} \tag{68}$$

There is also another hyperparameter, $\nu$, that acts as a thresholding parameter for the model, $\Xi$. During the training process, every $M$ epochs (500, in practice) all terms in $\Xi$ whose magnitude falls beneath $\nu$ will be pruned; only nonzero parameters in $\Xi$ will be updated in successive iterations.

In Section 4.2 of the Results, we use $\lambda_1 = 10^1, \lambda_2 = 0, \lambda_3 = 10^{-4}, \nu = 10^{-3}$. For Section 4.5 of the Results, starting from the pre-trained weights obtained in Section 4.2, we fix $\lambda_2$ and sweep a grid of $\lambda_1 \in \{10^{-5}, 10^{-4}, 10^{-3}\}$ and $\nu \in \{10^{-3}, 10^{-2}, 10^{-1}\}$, training each candidate for an additional 5,000 epochs (10,000 epochs for the neural population systems, to ensure adequate convergence given the more complex latent dynamics). The best model is chosen based on its minimum SLIC score. We do not tune $\lambda_2$ as: (i) we do not have access to $\dot{\mathbf{x}}$, and (ii) by enforcing both accurate reconstruction and consistent latent dynamics, we can also achieve the accurate reconstruction of future states that $\lambda_2$ is trying to achieve.

*Koopman-autoencoder (Koopman-AE)*. As in the previous two architectures, the Koopman-AE architecture [69] also uses a neural-network-based encoder($\varphi$)-decoder($\psi$) scheme. However, in addition to learning the weights for these functions, Koopman-AEs also attempt to learn a linear operator, $\mathbf{K}$, that steps the time-dependent latents, $\mathbf{z}(t)$, forward in time: $\mathbf{z}(t+1) = \mathbf{K}\mathbf{z}(t)$. Naturally, the first term in the loss function is a reconstruction term: $\mathcal{L}_{rec} = ||\mathbf{x} - \hat{\mathbf{x}}||_2^2$. As in SINDy-AEs, the other terms in the loss function consist of self-consistent forward projection in the latent space and ensuring that such a forward projection can faithfully forward project the input data. Specifically, there is $\mathcal{L}_{zf} = \sum_t \sum_{m=1}^{M} ||\mathbf{z}(t+m) - \mathbf{K}^m\mathbf{z}(t)||_2^2$ and $\mathcal{L}_{xf} = \sum_t \sum_{m=1}^{M} ||\mathbf{x}(t+m) - \psi(\mathbf{K}^m\mathbf{z}(t))||_2^2$. There is also a term penalizing large deviations in the encoder-decoder scheme: $\mathcal{L}_{ld} = ||\mathbf{x} - \hat{\mathbf{x}}||^6 + ||\mathbf{x}(t+1) - \psi(\mathbf{K}\mathbf{z}(t))||^6$; here $||\cdot||^6$ denotes the sum of errors to the sixth power. Finally, there is a regularization term for all trainable parameters (denoted $\mathbf{W}$): $\mathcal{L}_{reg} = ||\mathbf{W}||_2^2$. The total loss is written as:

$$\mathcal{L} = \lambda_1(\mathcal{L}_{rec} + \mathcal{L}_{xf}) + \mathcal{L}_{zf} + \lambda_2\mathcal{L}_{ld} + \lambda_3\mathcal{L}_{reg} \tag{69}$$

In Section 4.2 of the Results, we use $M = 3, \lambda_1 = 10^1, \lambda_2 = 10^{-7}, \lambda_3 = 10^{-14}$, as in [69]. In Section 4.5 of the Results, pre-trained Koopman-AE weights from Section 4.2 are used as the starting point. We fix $\lambda_2$ and perform a grid search over $\lambda_1 \in \{10^{-1}, 10^0, 10^1\}$ and $\lambda_3 \in \{10^{-14}, 10^{-7}, 10^{-1}\}$, training each candidate for an additional 5,000 epochs on 70% of the available timepoints of the new trajectory. The best model is selected as the one that minimizes the $L_2$ forecasting error on the held-out 30% of timepoints, after which only the Koopman matrix $\mathbf{K}$ is updated using the available prefix data.

For all autoencoder-based architectures (SINDy-AE and Koopman-AE) in Section 4.5, prefix adaptation uses 500 gradient epochs, except for the neural population systems which use 50 epochs; the neural systems use a larger hyperparameter search budget of 10,000 epochs per candidate (versus 5,000 for other systems) to ensure adequate convergence, and the shorter adaptation step reflects this larger initial investment. In systems where the input dimensionality changes between training and test conditions (e.g., the RNA–liposome system), the first encoder layer and final decoder layer are additionally updated to accommodate the new feature dimension, with all internal layers initialized from the pre-trained model.

## Nonlinear ODEs

The trajectories for the nonlinear dynamical systems in Section 2.2 were generated using either the `PredefinedDynamicalSystems.jl` package (in the case of the Lorenz, Rössler, and Guckenheimer-Holmes) or `DifferentialEquations.jl` (in the case of FitzHugh-Nagumo) in Julia. In each case, the `Tsit5()` integration method was used. The equations which govern each of the dynamical systems of study in this paper are provided below, along with default timespan and timestep; the default trajectory length referred to in Figure 2 is the range of the timespan over the timestep.

*Lorenz*

$$\frac{dx}{dt} = 10(y - x)$$
$$\frac{dy}{dt} = x(28 - z) - y$$
$$\frac{dz}{dt} = xy - \frac{8}{3}z$$

Default timespan: $t \in [0,10]$, timestep: 0.001.

*Rössler*

$$\frac{dx}{dt} = -y - z$$
$$\frac{dy}{dt} = x + 0.2y$$
$$\frac{dz}{dt} = 0.2 + z(x - 5.7)$$

Default timespan: $t \in [0,30]$, timestep: 0.01.

*FitzHugh-Nagumo*

$$\frac{dx}{dt} = x - \frac{x^3}{3} - y + 0.6$$
$$\frac{dy}{dt} = 0.08x + 0.056 - 0.064y$$

Default timespan: $t \in [0,80]$, timestep: 0.01.

*Guckenheimer-Holmes*

$$\frac{dx}{dt} = 0.4x - 20.25y + 3xz + 1.6z(x^2 + y^2)$$
$$\frac{dy}{dt} = 0.4y - 20.25x + 3yz$$
$$\frac{dz}{dt} = 1.7 - z^2 + 0.44(x^2 + y^2) - 0.4z^3$$

Default timespan: $t \in [0,20]$, timestep: 0.002.